\documentclass[12pt]{article}
\usepackage{amsmath}
\usepackage{graphicx}
\usepackage{enumerate}
\usepackage{natbib}
\usepackage{bm}
\usepackage[normalem]{ulem}

\newcommand{\blind}{0}

\usepackage[utf8]{inputenc}
\usepackage{algorithm,amssymb,amsthm,enumitem,hyphenat,microtype,xr-hyper}
\usepackage[dvipsnames]{xcolor}
\usepackage[colorlinks,allcolors=NavyBlue]{hyperref}
\usepackage{autonum} 

\newtheorem{theorem}{Theorem}
\newtheorem{proposition}[theorem]{Proposition}

\newtheorem*{example}{Example}

\newcommand{\ess}{\mathrm{ESS}}

\newcommand{\cL}{\mathcal{L}}

\newcommand{\bR}{\mathbb{R}}

\newcommand{\EE}{\mathbb{E}}
\newcommand{\cov}{\mathrm{cov}}
\newcommand{\Normal}{ \mathcal{N}}

\newcommand{\kl}[2]{\mathrm{KL}\BK{\, #1 \, 
\| \, #2\,}}
\newcommand{\distance}[3]{\mathrm{K}_{#1}\BK{\, #2 \, 
\| \, #3\,}}

\newcommand{\bra}[1]{\left< #1 \right>}
\newcommand{\BK}[1]{ {\left( #1 \right)} }
\newcommand{\sqBK}[1]{ {\left[ #1 \right]} }
\newcommand{\curBK}[1]{ {\left\{ #1 \right\}} }

\renewcommand{\phi}{\varphi}
\renewcommand{\epsilon}{\varepsilon}

\newcommand{\GG}{\mathrm{G}}
\newcommand{\Gg}{\mathrm{g}}  

\date{}

\begin{document}

\def\spacingset#1{\renewcommand{\baselinestretch}%
{#1}\small\normalsize} \spacingset{1}


\newcommand{\titleofarticle}{Doubly Adaptive Importance Sampling}

\if0\blind
{
  \title{\bf \titleofarticle}
  \author{Willem van den Boom\thanks{
  Email: \href{mailto:willem@wvdboom.nl}{willem@wvdboom.nl}.
    This work was partially supported by the Singapore Ministry of Education Academic Research Fund Tier~2 grant~MOE2016-T2-2-135.
    }\hspace{.2cm}\\
    Institute for Human Development and Potential, \\
    Agency for Science, Technology and Research, Singapore\\
    and \\
    Andrea Cremaschi \\
    School of Science and Technology,
    IE University, Madrid, Spain \\
    and \\
    Alexandre H.\ Thiery \\
    Department of Statistics and Data Science, \\
    National University of Singapore}
  \maketitle
} \fi

\if1\blind
{
  \bigskip
  \bigskip
  \bigskip
  \begin{center}
    {\LARGE\bf \titleofarticle}
\end{center}
  \medskip
} \fi

\bigskip
\begin{abstract} 
We propose an adaptive importance sampling scheme for Gaussian approximations of intractable posteriors. Optimization\hyp{}based approximations like variational inference can be too inaccurate while existing Monte Carlo methods can be too slow. Therefore, we propose a hybrid where, at each iteration, the Monte Carlo effective sample size can be guaranteed at a fixed computational cost by interpolating between natural\hyp{}gradient variational inference and importance sampling. The amount of damping in the updates adapts to the posterior and guarantees the effective sample size. Gaussianity enables the use of Stein's lemma to obtain gradient\hyp{}based optimization in the highly damped variational inference regime and a reduction of Monte Carlo error for undamped adaptive importance sampling. The result is a generic, embarrassingly parallel and adaptive posterior approximation method. Numerical studies on simulated and real data show its competitiveness with other, less general methods.
\end{abstract}

\noindent%
{\it Keywords:} 
Adaptive Monte Carlo;
Embarrassingly parallel computing;
Gaussian posterior approximation;
Natural\hyp{}gradient variational inference;
Stein's lemma
\vfill

\newpage
\spacingset{1.5} 

\section{Introduction}
\label{sec:intro}
It is common in several modeling settings to encounter distributions that are only known up to proportionality. Some examples are models characterized by an intractable likelihood, complex prior specifications or general non-conjugate Bayesian models.
The goal of this work is to propose a novel approach to approximate an $\bR^d$-valued probability distribution $\pi$ for which the normalizing constant is unknown in closed form. Although the distribution $\pi$ can be approximated by numerical integration in low-dimensional settings, the problem is typically challenging in higher dimensions due to the curse of dimensionality. Popular approximation schemes include optimization-based methods such as Laplace approximation, variational inference \citep[VI,][]{Blei2017} and expectation propagation \citep[EP,][]{Minka2001}, as well as  Monte Carlo (MC) approaches such as Markov chain MC, (adaptive) importance sampling \citep[IS,][]{Bugallo2017} or sequential Monte Carlo \citep[SMC,][]{DelMoral2006}. However, optimization\hyp{}based approximations are often too inaccurate while MC can be computationally too expensive. This motivates hybrid approaches that blend features of optimization and MC methods \citep{Jerfel2021}. We propose a novel hybrid method that adaptively interpolates between the computational speed and approximate nature of VI \citep{Khan2018} and the accuracy and computational cost of IS.

The task of approximating a density known only up to proportionality typically arises in Bayesian statistics, where intractable posterior distributions are commonly encountered. Let  $x\in\bR^d$ indicate a $d$-dimensional parameter vector and $y\in\bR^n$ be the data, so that $\ell(y\!\mid\! x)$ provides the model likelihood. Let $p_0$ be the prior distribution of the parameter $x$ and $p_y$ the marginal density of $y$. Then, by Bayes' rule, the posterior distribution of $x$ given $y$ is obtained as $\pi(x) = \ell(y\!\mid\! x)\, p_0(x) / p_y(y)$, where we indicate the distributions and the corresponding densities with the same symbols. The normalizing constant $p_y(y)$ of the posterior density $\pi$ is often intractable and cumbersome to approximate in high-dimensional settings, such as Bayesian inverse problems \citep{Stuart2010}.

We devise an adaptive IS (AIS) scheme that iteratively adapts a proposal distribution $q$ by matching its sufficient statistics
with
an annealed version $q_\gamma$ of the target distribution $\pi$.
Specifically,
$q_\gamma$ is obtained through an
adaptive
\emph{damping}\footnote{We borrow the term `damping' from work on expectation propagation \citep[Section~5.2]{Vehtari2020}.}
mechanism that guarantees a target effective sample size (ESS) used for quantifying the MC error.
In particular, $q_\gamma(x)\propto q^{1-\gamma}(x)\,\pi^\gamma(x)$ where the damping parameter $\gamma$, which specifies the annealing at each iteration of the algorithm, is obtained by numerically solving a fixed lower bound on the ESS.
The adaptation of both the proposal $q$ and the damping $\gamma$ motivates the name \emph{doubly adaptive importance sampling} (DAIS). The proposed approach is based on IS and can consequently easily leverage parallel computations and modern compute environments.
We highlight two contributions to the AIS literature:
\begin{enumerate}
\item
Differently from previous AIS approaches, we allow the iterative scheme to converge to (and thus terminate at) an annealed target distribution were the amount of annealing (if any) depends on the ESS bound, sample size and shape of the target.
\item
Working with Gaussian proposals,\footnote{See Section~\ref{sec:discussion}
for adapting DAIS to 
more general proposals.}
we use Stein's lemma to build a variance reduction scheme that significantly enhances the robustness of the proposed method.
\end{enumerate}

We set the proposed methodology in the context of variational inference (VI). In its most common form, VI determines an approximating distribution $q$ by minimizing the Kullback-Leibler (KL) divergence $\kl{q}{\pi}$ of $\pi$ from $q$ over a tractable family of distributions. The quantity $\kl{q}{\pi}$, often referred to as the \emph{reverse} KL divergence, involves an expectation with respect to the tractable approximating distribution $q$. While it is relatively straightforward to implement VI, the use of the reverse KL divergence is known to often yield an approximating distribution $q$ with lower variance than $\pi$ \citep{Minka2005,Li2016,Jerfel2021} and thus overconfident Bayesian inference. Minimizing the \emph{forward} KL divergence $\kl{\pi}{q}$ can mitigate these issues but is typically computationally challenging since it involves an expectation with respect to the intractable distribution $\pi$.
Linking both extremes, the $\alpha$-divergence $\distance{\alpha}{\pi}{q}$ \citep{
Minka2005,Li2016} interpolates between the reverse and the forward KL divergences using the parameter $\alpha$.

Recall that damping is based on ESS, which closely relates to the $\chi^2$\hyp{}divergence and thus MC error \citep{SanzAlonso2018}.
Still, we find that the fixed points of DAIS' iterative procedure can be described as a stationary point of the functional $q \mapsto \distance{\alpha}{\pi}{q}$ for a parameter $0<\alpha<1$ related to the amount of damping used within our method.
The proposed adaptive damping scheme thus produces an automatic trade-off between minimizing the computationally more convenient reverse KL divergence and the forward KL divergence, which yields more accurate approximations. Finally, we establish that, in the limit of maximally damped updates (i.e.\ slow updates), our method corresponds to minimizing the reverse KL divergence $\kl{q}{\pi}$ with a natural\hyp{}gradient descent scheme.

The rest of the article is organized as follows.
Section~\ref{sec:DAIS} introduces DAIS.
Section~\ref{sec:related} discusses related work.
Section~\ref{sec:theory} analyses the link with natural\hyp{}gradient VI and $\alpha$-divergence. 
Section~\ref{sec:examples} presents empirical results.
Section~\ref{sec:discussion} concludes.

\section{Doubly Adaptive Importance Sampling}
\label{sec:DAIS}

\subsection{Algorithm Setting and Notation}
\label{sec:assum}

We consider a target distribution
on $\bR^d$ with strictly positive and continuously differentiable density $\pi$
with respect to the Lebesgue measure.
We constrain ourselves to building Gaussian approximations of $\pi$, although parts of our development also apply to other families of approximating distributions, as discussed in Section~\ref{sec:discussion}. DAIS iteratively builds a sequence of Gaussian approximations to the target distribution $\pi$. At iteration $t \geq 1$, the current Gaussian approximation is denoted by $q_t(x) \equiv \Normal\left(x\!\mid\!\mu_t, \Gamma_t\right)$ for a mean vector $\mu_t \in \bR^d$ and positive-definite covariance matrix $\Gamma_t$.
DAIS requires the log-target density $\log \pi(x)$ to be known up to an additive constant and its gradient $\nabla\!_x \log \pi(x)$ to be efficiently evaluated:
if a computer program exists for the evaluation of $\log \pi(x)$,
then algorithmic differentiation can provide its gradient at a similar computational cost
as evaluation of the target density \citep[Section~3.3]{Griewank2008}.

Throughout the article, we denote expectations with respect to a distribution $\eta$ by $\EE_\eta[\phi(X)] = \int \phi(x)\, \eta(x)\,dx$. For two vectors $u,v \in \bR^d$, we indicate their inner product by $\bra{u,v} = \sum\limits_{i=1}^d u_i v_i$, and their outer product by $u \otimes v \equiv u \, v^\top \in \bR^{d\times d}$. For two vector-valued functions $U: \bR^d \to \bR^{d_U}$ and $V: \bR^d \to \bR^{d_V}$, and an $\bR^d$-valued random variable $X$ with distribution $\eta$, the covariance matrix between the random variables $U(X)$ and $V(X)$ is denoted as $\cov_\eta[U(X), V(X)] \in \bR^{d_U\times d_V}$.
We denote the covariance matrix by $\cov_\eta[X] = \cov_\eta[X,X]$.
Finally, the KL divergence between two
densities
$p \ll q$ is defined as $\kl{p}{q} = \EE_p[\log\{ p(X)/q(X)\}]$.

\subsection{Adaptive Gaussian Approximations}
\label{sec:damping}

We can assess the quality of the approximation to the target distribution $\pi(x)$ by measuring the closeness of the two probability distributions. This can be done in different ways, such as minimizing the forward or reverse KL divergences. These divergences measure closeness differently.
DAIS builds a Gaussian approximation $q_\star$ to the target distribution whose first two moments are matched to those of $\pi$. This corresponds to minimizing the forward KL \citep[Equation~(10.187)]{Bishop2006},
\begin{align}
q_\star \; = \; \mathrm{argmin} \; \curBK{ \kl{\pi}{q} \; : \; q \in \mathcal{Q}_{\mathrm{gauss}}},
\end{align}
where $\mathcal{Q}_{\mathrm{gauss}}$ denotes the exponential family of Gaussian distributions. This objective is desirable in a Bayesian setting as the posterior mean and covariance are often of interest.

The DAIS procedure is initialized from a user-specified Gaussian approximation $q_1(x) \equiv \Normal\left(x\!\mid\!\mu_1, \Gamma_1\right)$.
Approaches for setting the initial approximation include using (a Gaussian approximation to) the prior distribution or a Laplace approximation to the target distribution $\pi$, which we use in Section~\ref{sec:logistic}. Also, more sophisticated procedures are possible such as running DAIS multiple times with random initializations.
The quality of the initial approximation can greatly influence the number of iterations required for convergence and might affect quality of the final approximation, e.g.\ if $\pi$ is multimodal.

Given the current approximation $q_t$, an improved Gaussian approximation $q_{t+1}$ is obtained by matching its moments with an annealed distribution $q_{t, \gamma_t}$ that interpolates between the current Gaussian distribution $q_t$ and the target distribution $\pi$.
The annealing is used since
directly targeting $\pi$ might result in too high MC error of the moment estimates if $q_t$ is too different from $\pi$.
The intermediate target, defined as $q_{t,\gamma_t}(x) \propto q_t^{1-\gamma_t}(x)\, \pi^{\gamma_t}(x)$, is coined the \emph{damped target density} in analogy with damped expectation propagation (EP) \citep[Section~5.2]{Vehtari2020}. The parameter $0 < \gamma_t \leq 1$ is referred to as the \emph{damping parameter}
with smaller $\gamma_t$ corresponding to more damping of the update from $q_t$ to $q_{t+1}$.
Furthermore,
\begin{align} \label{eq:damping}
q_{t,\gamma_t}(x) \propto q_t(x)\, e^{\gamma_t \Phi_t(x)}
\qquad \textrm{for} \qquad
\Phi_t(x) = \log \pi(x) - \log q_t(x).
\end{align}
The function $\Phi_t: \bR^d \to \bR$ captures the discrepancy between the current approximation $q_t$ and the target distribution $\pi$. Since DAIS only requires the gradient of $\Phi_t$, the target distribution can be specified up to a multiplicative constant.

The mean $\mu_{t, \gamma_t}$ and covariance $\Gamma_{t, \gamma_t}$ of the damped target density $q_{t, \gamma_t}$ can be expressed as perturbations of the current parameters $\mu_t$ and $\Gamma_t$:
%
%
\begin{align} \label{eq.CV.mean.cov}
\left\{
\begin{aligned}
\mu_{t, \gamma_t} 
&=
\EE_{q_{t,\gamma_t}}[X]
=
\mu_t + \gamma_t \, \Gg_\mu(q_{t,\gamma_t})\\
\Gamma_{t, \gamma_t} 
&=
\cov_{q_{t,\gamma_t}}[X]
=
\Gamma_t + \gamma_t \, \GG_\Gamma(q_{t,\gamma_t})
\end{aligned}
\right.
\quad \text{where} \quad
\left\{
\begin{aligned}
\Gg_\mu(q_{t,\gamma_t}) 
&= \EE_{q_{t, \gamma_t}}[\Gamma_t \nabla \Phi_t(X)]
\\
\GG_\Gamma(q_{t,\gamma_t})
&= \cov_{q_{t, \gamma_t}}[  \Gamma_t \nabla \Phi_t(X), X].
\end{aligned}
\right.
\end{align}
The proof of the identities in \eqref{eq.CV.mean.cov} is presented in Section~\ref{sec.CV} and Appendix~\ref{ap:Stein}.
Furthermore, Section~\ref{sec:theory} connects the quantities $\Gg_\mu(q_{t, \gamma_t})$ and $\GG_\Gamma(q_{t, \gamma_t})$ to the (negative) natural gradients of both the functionals $q_t \mapsto \kl{q_t}{\pi}$ and $q_t \mapsto \kl{\pi}{q_t}$. 
Since both $\Gg_\mu(q_{t, \gamma_t})$ and $\GG_\Gamma(q_{t, \gamma_t})$ are expressed as expectations with respect to the damped target density $q_{t, \gamma_t}$, these quantities can be estimated as $\widehat{\Gg}_\mu(q_{t, \gamma_t})$ and $\widehat{\GG}_\Gamma(q_{t, \gamma_t})$ with self-normalized IS with $S \geq 1$ samples generated from $q_t$.

The damping parameter $0 < \gamma_t \leq 1$, that controls the closeness of the intermediate distribution $q_{t, \gamma_t}$ to the target distribution $\pi$, is chosen adaptively so that the ESS is above a user-specified threshold $1 < N_{\ess} < S$.
Here, we condition on
$S$ samples $x_{t,1:S} = (x_{t,1}, \ldots, x_{t,S})$ from the proposal distribution $q_t$.
Then, the associated ESS is computed as:
%
%
\begin{align}
\ess\{ w_{t,1:S}(\gamma) \} = \frac{ \big\{ \sum_s w_{t,s}(\gamma) \big\}^2 }{ \sum_s w_{t,s}^2(\gamma) }
\qquad \text{with} \qquad
w_{t,s}(\gamma) = \exp\{\gamma\,\Phi_t(x_s)\} \propto \frac{q_{t,\gamma}(x_{t,s})}{ q_{t}(x_{t,s})}.
\end{align}
Since $\gamma \mapsto ESS\{ w_{t,1:S}(\gamma) \}$ is
a continuous and decreasing function of $0 < \gamma \leq 1$ \citep[Lemma~3.1]{Beskos2016}, then the least amount of damping
given the particles $x_{t,1:S}$
at iteration $t$ can efficiently be computed with a standard root-finding method such as the bisection method, solving:
\begin{align+}  
\label{eq:epsilon_t}    
\gamma_t = \max \Big\{ \gamma \in (0,1] \; : \; \ess\{ w_{t,1:S}(\gamma) \} \geq N_{\ess} \Big\}.
\end{align+}
The resulting scheme is summarized in Algorithm~\ref{alg:DAIS}.

\begin{algorithm}[tb]

	\caption{Doubly adaptive importance sampling} \label{alg:DAIS}
	
\begin{enumerate}
\item
    Initialize the algorithm with $q_1(x) \equiv \Normal(x\!\mid\!\mu_1, \Gamma_1)$.
\item
    For $t=1,2,\dots$ until the ELBO no longer improves:
    \begin{enumerate}
        \item
        Sample $x_{t,s} \sim q_t(x) \equiv \Normal\left(x\!\mid\!\mu_t, \Gamma_t\right)$ independently for $s=1,\dots,S$.
        \item \label{step:S_eff}
        For a damping parameter $\gamma$, the unnormalized importance weights are:\\
        $
        w_{t,s}(\gamma) = \exp\{\gamma\, \Phi_t(x_{t,s})\}
         \propto q_{t,\gamma}(x_{t,s}) / q_{t}(x_{t,s})
        $\\
        and the ESS is then:\\
$
\ess_{t}(\gamma)
=
\big\{ \sum_{s = 1}^S w_{t,s}(\gamma) \big\}^2 /
\sum_{s = 1}^S w_{t,s}^2(\gamma)
$

For a given threshold $1<N_{\text{ESS}}<S$, update the damping parameter $\gamma$:

$
\begin{array}{ll}
    \gamma_t = 1, & \text{if } \; \ess_{t}(1) \geq N_{\ess} \\
    \gamma_t = \max \left\{\gamma\in(0,1) : \ess_{t}(\gamma) \geq N_{\ess}\right\}, & \text{if } \; \ess_{t}(1) < N_{\ess}
\end{array}
$
        \item
        Compute $q_{t+1}(x)\equiv \Normal(x\!\mid\!\mu_{t+1}, \Gamma_{t+1})$ using the samples $x_{t,s}$ weighted by $w_{t,s}(\gamma_t)$:
        set $\mu_{t+1}$ and $\Gamma_{t+1}$ equal to an IS estimate of the right-hand sides of \eqref{eq.CV.mean.cov}.
    \end{enumerate}
    \item
    Return the final Gaussian $q_{t+1}$ as the approximation to the target distribution $\pi$.
\end{enumerate}

\end{algorithm}

As explained in Section \ref{sec:theory}, the quantities $\Gg_\mu(q_{t, \gamma_t})$ and $\GG_\Gamma(q_{t, \gamma_t})$ can heuristically be thought of as (natural) gradients. This motivates the updates
\begin{align}
\left\{
\begin{aligned}
\mu_{t+1} &= \mu_t + \zeta_t \, \widehat{\Gg}_\mu(q_{t, \gamma_t}) \\
\Gamma_{t+1} &= \Gamma_t + \zeta_t \, \widehat{\GG}_\Gamma(q_{t, \gamma_t})
\end{aligned}
\right.
\end{align}
for a sequence of learning rates $\zeta_t > 0$. The choice $\zeta_t = \gamma_t$ corresponds to matching the first two moments of $q_{t+1}$ to the (estimate of) the first two moments of the damped target distribution $q_{t, \gamma_t}$. Since $\widehat{\Gg}_\mu(q_{t, \gamma_t})$ and $\widehat{\GG}_\Gamma(q_{t, \gamma_t})$ are only stochastic estimates, for improved stability, we advocate choosing $\zeta_t = c \, \gamma_t$ for a \emph{robustness parameter} $0<c\leq 1$.
We monitor convergence
and decide when to terminate Algorithm~\ref{alg:DAIS}
using the Evidence Lower BOund (ELBO)
as discussed in Appendix~\ref{ap:monitoring}.

\subsection{Control Variate for Gaussian Perturbations}
\label{sec.CV}

\begin{figure}[tb]
\begin{center}
\includegraphics[width=0.49\textwidth]{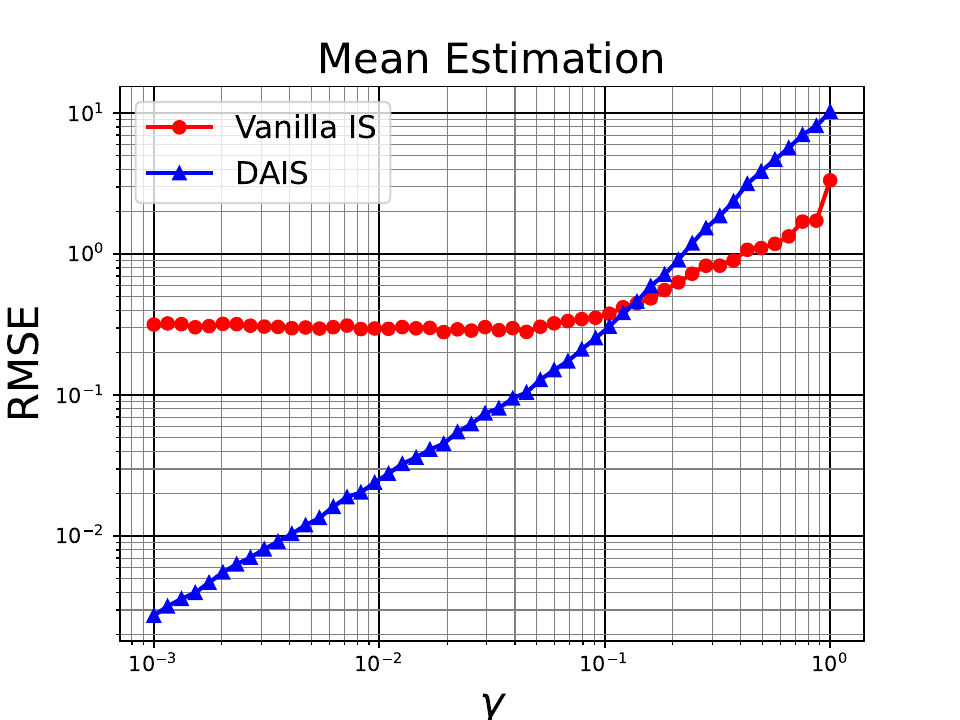}
\includegraphics[width=0.49\textwidth]{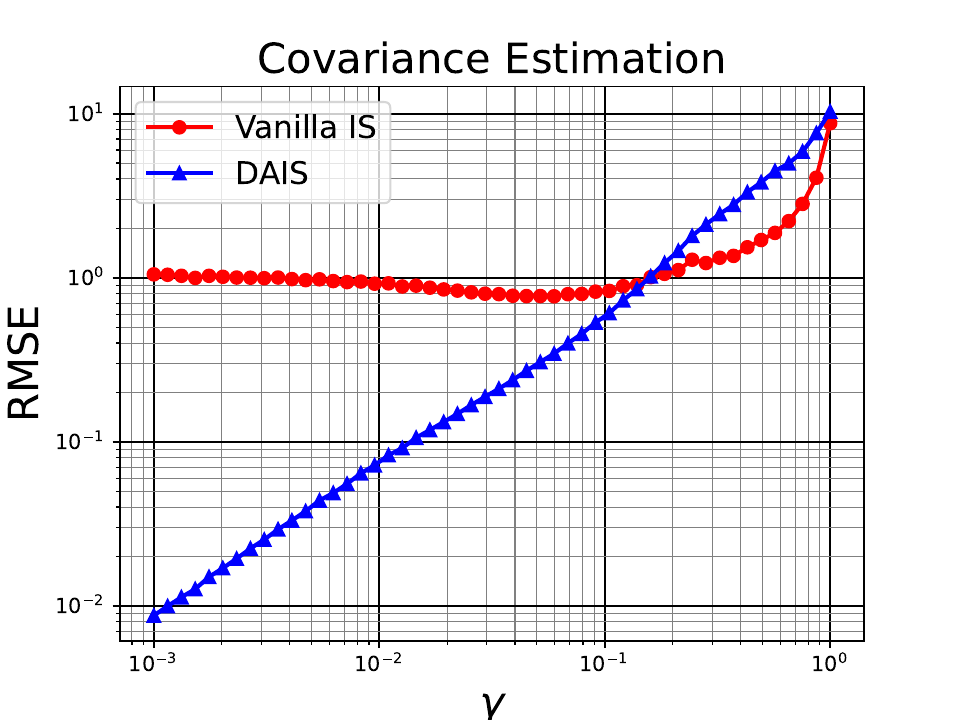}
\end{center}
\caption{Estimation of $\widehat{\mu}_{t, \gamma_t}$ and $\widehat{\Gamma}_{t, \gamma_t}$ as a function of $\gamma_t$ with standard self-normalized IS and with DAIS based on \eqref{eq.CV.mean.cov} with $S=10^2$ particles. The proposal $q_t(x) = \Normal(x\!\mid\! 0,\mathbf{1}_D)$ is a standard isotropic distribution in dimension $D=10$ and the target distribution is also Gaussian $\pi(x) = \Normal(x\!\mid\! m, \Sigma)$ with mean $m = (1, \ldots, 1) \in \bR^D$ and covariance $\Sigma$ with $\Sigma_{i,j} = 0.9 + 0.1 \, \delta(i=j)$.
}
\label{fig:control_variates}
\end{figure}

Since $q_{t, \gamma_t}$ is a perturbation of the current Gaussian approximation $q_t$, estimating the moments $\mu_{t, \gamma_t}$ and $\Gamma_{t, \gamma_t}$ from scratch is statistically suboptimal.
Instead, we derive update equations using \citeauthor{stein1972bound}'s \citeyearpar{stein1972bound} identity: for a probability density $p(x)$ on $\bR^d$ and a continuously differentiable test function $\phi: \bR^d \to \bR^d$, we have that \citep[Proposition~2]{Oates2017}
\begin{align} \label{eq.stein}
\EE_p \sqBK{ \BK{\bra{\nabla \log p, \phi} + \mathrm{div}  \phi }(X) } = 0
\end{align}
Equation~\eqref{eq.stein} follows from an integration by parts that is justified under mild growth and regularity assumptions \citep{mira2013zero,Oates2017}.
As derived in Appendix~\ref{ap:Stein}, applying \eqref{eq.stein} to the annealed density $q_{t, \gamma_t}$ and appropriate test functions gives
Equation~\eqref{eq.CV.mean.cov}.\footnote{A similar use of Stein's lemma for a gradient flow in VI appears in Section~4.2.2 of \citet{Chen2024}.}
The identities in \eqref{eq.CV.mean.cov} show that, given knowledge of $\mu_t$ and $\Gamma_t$, the first two moments of $q_{t, \gamma_t}$ can be estimated with a root-mean-square error (RMSE) of order $\gamma_t / \sqrt{S}$
when using $S$ samples.
A standard IS procedure that estimates these two quantities from scratch (i.e.\ without exploiting knowledge of $\mu_t$ and $\Gamma_t$) would typically lead to an RMSE of order $1/\sqrt{S}$,
i.e.\ the MC error does not vanish as $\gamma_t \to 0$.
Specifically,
using results and regularity conditions from
\citet{Agapiou2017},
we have the following (see Appendix~\ref{ap:rmse} for a proof).
\begin{proposition} \label{prop:rmse}
Assume that
$\EE_{q_t}[e^{2\gamma_t \Phi_t(X)}]<\infty$.
\begin{enumerate}[label=(\roman*), font=\upshape]
\item
If
$\EE_{q_t}[X_i^2\, X_j^2\, e^{2\gamma_t \Phi_t(X)}]<\infty$ for all $i,j$,
then
the self-normalized IS estimators based on the standard moments in the left-hand side of \eqref{eq.CV.mean.cov} have an RMSE of order $1/\sqrt{S}$ as $S\to\infty$.
\item
If
$\EE_{q_t}[\{\nabla_i\Phi_t(X)\}^2\, e^{2\gamma_t \Phi_t(X)}]<\infty$
and
$\EE_{q_t}[\{\nabla_i\Phi_t(X)\}^2 X_j^2\, e^{2\gamma_t \Phi_t(X)}]<\infty$
for all $i,j$,
then
the self-normalized IS estimators based on the gradient-based expressions in the right-hand side of \eqref{eq.CV.mean.cov} have an RMSE of order $\gamma_t/\sqrt{S}$ as $S\to\infty$.
\end{enumerate}
\end{proposition}
\noindent
Furthermore, when $q_t$ is a good approximation to the target distribution $\pi$, which is expected as the DAIS procedure progresses, the discrepancy function $\Phi_t$ and its gradient typically become small, leading to improved robustness.
We conclude this section by illustrating the statistical advantages of estimating the first two moments $\mu_{t, \gamma_t}$ and $\Gamma_{t, \gamma_t}$ of $q_{t, \gamma_t}$ through \eqref{eq.CV.mean.cov} when compared to a naive IS estimation of these two quantities. 
\begin{example}
Consider a tractable setting where $q_t(x) = \Normal(x\!\mid\! 0,\mathrm{I}_d)$ is a standard isotropic distribution of dimension $d = 10$ and the target distribution is also Gaussian $\pi(x) = \Normal\left(x\!\mid\! m, \Sigma\right)$ with mean $m = (1, \ldots, 1)$ and covariance $\Sigma$ with $\Sigma_{i,j} = 0.9 + 0.1 \, {\delta(i=j)}$.
Figure \ref{fig:control_variates} reports, as a function of $\gamma_t$, the RMSE quantities $\EE[\|\widehat{\mu}_{t, \gamma_t} - \mu_{t, \gamma_t}\|^2]^{1/2}$ and ${\EE[\| \widehat{\Gamma}_{t, \gamma_t} - \Gamma_{t, \gamma_t} \|_{\mathrm{F}}^2]^{1/2}}$, where $\|M\|_{\mathrm{F}}^2 = \sum M_{i,j}^2$ is the squared Frobenius norm of the matrix $M$. The RMSEs are approximated with $10^2$ independent experiments and the IS estimates use $S=10^2$ particles.
\end{example}

%
%

\section{Related Work}
\label{sec:related}

We now provide an overview of existing posterior approximation approaches to contextualize the proposed approach DAIS and highlight relevant connections.

\subsection{Methods Based on Importance Sampling}
\label{sec:ais}

We start by reviewing AIS \citep{Bugallo2017} with Gaussian proposal distributions to approximate the target $\pi$. 
AIS is firstly initialized to $q_1$, a Gaussian distribution with mean $\mu_1$ and covariance $\Gamma_1$, for instance obtained from the prior distribution or from an initial approximation to the target $\pi$. Then, the proposal $q_{t+1}$ at each iteration is determined via the IS approximation to $\pi$ from the previous iteration. That is,
$\mu_{t+1}$ and $\Gamma_{t+1}$ are chosen so that $q_{t+1}$ is closer to $\pi$, resulting in a more accurate IS approximation.

AIS \citep{Bugallo2017} and DAIS iteratively improve $q_t$ via moment matching.
A novelty of DAIS is in the choice of the new parameters $\mu_{t+1}$ and $\Gamma_{t+1}$ in \eqref{eq.CV.mean.cov}, which is based on an application of Stein's lemma.
Another difference from
previous AIS schemes is the adaptation of the IS target in the ultimate iteration to guarantee ESS.
\citet{Ryu2015},
\citet{Schuster2015}, \citet{Akyildiz2021}, \citet{Elvira2022b}, and \citet{Elvira2023}
propose AIS approaches that, like ours, use the gradient of the target $\pi$ to update the proposal.
Though, their gradients appear as part of optimization algorithms
while ours derives from moment matching with Stein's lemma. 

Smoothing of IS weights, of which the damping in DAIS is a special case, has been employed in IS \citep[e.g.][]{Koblents2013,Vehtari2021} and in AIS \citep{Paananen2021}. Most similarly to DAIS, \citet{Koblents2013} base their decision whether to temper the weights on ESS, though they do not adapt the amount of tempering to the target.
Like DAIS but with a different smoothing method, \citet{Paananen2021} use a stopping criterion based on the regularity of the weights to determine the number of AIS iterations.
Their proposal distributions, arising from specific tasks such as Bayesian cross-validation, are complicated while DAIS considers Gaussian proposals with a focus on approximating the posterior mean and covariance of the target distribution.
As we do in Section~\ref{sec:theory},
\citet{Guilmeau2024}
link an AIS scheme to $\alpha$-divergence minimization.
There, $\alpha$ is adapted through a one-to-one correspondence with the tail behaviour of the IS proposal distribution.

Similarly to DAIS, SMC \citep{DelMoral2006} and annealed IS \citep{Neal2001} adapt the target and proposal across iterations. Moreover, automatic tempering using ESS is also used in adaptive SMC \citep[e.g.][Algorithm~17.3]{Chopin2020}. In these methods, proposals are discrete distributions based on reweighted, resampled or rejuvenated particles while DAIS adapts a Gaussian proposal.
Typically, proposal adaptation is based on geometric averaging which coincides with the damping in \eqref{eq:damping}
and entropic mirror descent on $\kl{q_t}{\pi}$ \citep{Chopin2024}, though moment matching has been explored as well \citep{Grosse2013}.
DAIS uses both damping and moment matching simultaneously.

\subsection{Optimization-based Methods}
\label{sec:related_optim}

Moment matching is fundamental to EP \citep{Minka2001} and expectation consistent approximate inference \citep{Opper2005}. These methods usually entail such matching iteratively across factors of the target density $\pi$. That is, expectations are propagated across a Bayesian network.
This contrasts with our matching, which uses all of $\pi$ at once, e.g. $\mu_{t+1} = \EE_{q_{t, \gamma_t}}[X]$. Disregarding this discrepancy, the damped moment matching of DAIS is equivalent to damping in EP \citep[Section~5.2]{Vehtari2020} and to the $\alpha$-divergence minimization scheme in Equation~(18) of \citet{Minka2005}. Like DAIS, the EP methods by \citet{Wiegerinck2003}, \citet{Minka2004}
and \citet{Hernandez-Lobato2016}
minimize the $\alpha$-divergence.
The updates in \eqref{eq.CV.mean.cov} involve taking the expectation, or smoothing, of gradients. \citet{Dehaene2016} links smoothed gradients to EP and minimization of $\alpha$-divergence.

\citet{Prangle2022} consider the same damped target $q_{t, \gamma_t}$, also based on ESS, while iteratively updating an IS proposal $q_t$ as in DAIS.
Differences include that $q_t$ is not a Gaussian distribution but a normalizing flow
and that $q_t$ is updated via gradient descent for the objective $\kl{ q_{t, \gamma_t} }{ q_{t+1} }$.
DAIS updates $q_t$ through moment matching, e.g. $\mu_{t+1} = \EE_{q_{t, \gamma_t}}[X]$,
which directly targets the minimizer of $\kl{ q_{t, \gamma_t} }{ q_{t+1} }$.

There is a VI literature on improving variational objectives and approximating families via MC
\citep[e.g.][]{Li2016,Ruiz2019} including IS
\citep[e.g.][]{Domke2018,Wang2018} and AIS \citep{Han2017,Jerfel2021}.
DAIS constitutes a substantially different hybrid between VI and MC as it performs AIS that happens to recover natural\hyp{}gradient VI via damping as $\gamma_t\to 0$ (see Section~\ref{sec:theory}).
Some VI methods \citep[e.g.][]{Li2016} replace the reverse KL divergence by $\alpha$-divergence,
which DAIS effectively also minimizes (see Section~\ref{sec:theory}).
Moreover, \citet{Daudel2023} and \citet{Guilmeau2024b} obtain damped moment matching updates as in DAIS.
However, DAIS, derived as AIS instead of a change in VI objective, yields principled adaption of the damping parameter $\alpha=\gamma_t$. In this context, \citet{Wang2018} consider adaptation of the divergence based on tail probabilities of importance weights. \citet{Yao2018} evaluate the accuracy of VI using IS.
\citet{Liu2016} use Stein's lemma for VI.
They minimize the reverse KL divergence
via a functional gradient descent derived from the Stein discrepancy. \citet{Han2017} expand that Stein VI method to use AIS.

\section{Analysis of DAIS}
\label{sec:theory}

Implementing the DAIS algorithm with learning rate $\zeta_t = \gamma_t$ corresponds to iteratively matching, up to MC variability, the first two moments of the damped target density $q_{t, \gamma}$ to the ones of the next Gaussian approximation $q_{t+1}$. This demonstrates that the algorithm does not depend on the mean-covariance parametrization of the Gaussian family, and that any other parametrization would lead to exactly the same sequence of Gaussian approximations.
This remark motivates the connections described in this section between DAIS and the natural\hyp{}gradient descent method \citep{amari1998natural}.
We defer derivations to Appendix~\ref{ap:theory}.

While the standard gradient is the steepest descent direction when the usual Euclidean distance is used, the natural gradient is the steepest descent direction in the space of distributions where distance is measured by the KL divergence \citep{Martens20}. In particular, natural\hyp{}gradient flows are parametrization invariant. In the Gaussian setting of this article, the natural\hyp{}gradient flow for minimizing a loss function $\cL(\mu, \Gamma)$ over the space of Gaussian distributions $q \in \mathcal{Q}_{\mathrm{gauss}}$ is 
\begin{align+}  
\label{eq.NG.descent}
\left\{
\begin{aligned}
\frac{d \mu}{dt} &= - \Gamma \, \nabla_{\mu} \cL\\
\frac{d \Gamma^{-1}}{dt} &= 2 \, \nabla_{\Gamma} \cL\\
\end{aligned}
\right.
\qquad \Longleftrightarrow \qquad
\left\{
\begin{aligned}
\frac{d \mu}{dt} &= - \Gamma \, \nabla_{\!\mu} \cL \equiv -\widetilde{\nabla}_{\!\mu} \cL\\
\frac{d \Gamma}{dt} &= -2 \, \Gamma \BK{\nabla_{\Gamma} \cL} \Gamma
\equiv -\widetilde{\nabla}_{\Gamma} \cL,
\end{aligned}
\right.
\end{align+}
where $\widetilde{\nabla}_{\mu} \cL = \Gamma \, \nabla_{\mu} \cL$ and $\widetilde{\nabla}_{\Gamma} \cL = 2 \, \Gamma \BK{\nabla_{\Gamma} \cL} \Gamma$ denote the natural gradient with respect to the mean and covariance parameters, respectively.
For $\Gg_\mu(q_{t, \gamma_t})$ and $\GG_\Gamma(q_{t, \gamma_t})$ defined in \eqref{eq.CV.mean.cov},
\begin{align}
\label{eq.NG.rev}
\left\{
\begin{aligned}
\lim_{\gamma_t \to 0} 
\Gg_\mu(q_{t, \gamma_t}) &= \Gamma \EE_{q_t}[ \, \nabla_x \log \pi(X)] = - \widetilde{\nabla}_{\mu} \kl{q_t}{\pi} \\
\lim_{\gamma_t \to 0} 
\GG_\Gamma(q_{t, \gamma_t}) & = \Gamma\,\EE_{q_t}[\nabla^2_{xx} \log \pi(X)]\,\Gamma + \Gamma = -\widetilde{\nabla}_{\Gamma} \kl{q_t}{\pi}.
\end{aligned}
\right.
\end{align}
Equation~\eqref{eq.NG.rev} shows that, in the limit of small damping parameter $\gamma_t \to 0$, DAIS can be understood as a natural\hyp{}gradient descent for minimizing the reverse KL. Furthermore,
\begin{align}
\label{eq.NG.fwd}
\left\{
\begin{aligned}
\lim_{\gamma_t \to 1} 
\Gg_\mu(q_{t, \gamma_t}) &= \mu_{\pi} - \mu = - \widetilde{\nabla}_{\mu} \kl{\pi}{q_t} \\
\lim_{\gamma_t \to 1} 
\GG_\Gamma(q_{t, \gamma_t}) & = \Gamma_{\pi} - \Gamma
= - \widetilde{\nabla}_{\Gamma} \kl{\pi}{q_t} - (\mu_{\pi}-\mu) \otimes (\mu_{\pi}-\mu)
\end{aligned}
\right.
\end{align}
where $\mu_{\pi} = \EE_{\pi}[X]$ and $\Gamma_{\pi} = \cov_{\pi}[X]$.
Equation~\eqref{eq.NG.fwd} establishes a connection between DAIS and the natural\hyp{}gradient flow for minimizing the forward KL, whose global minimizer is indeed given by the Gaussian distribution with first two moments matching those of the target distribution $\pi$.

To conclude this section, we characterize the limiting distribution obtained by the DAIS methodology. For this purpose, assume that the DAIS algorithm has converged towards an approximating distribution $q_\infty(x) = \Normal(x\!\mid\!\mu_\infty, \Gamma_\infty)$ with final damping parameter $0<\gamma_{\infty}<1$. The moment matching conditions mean that
\begin{align} \label{eq.alpha.div}
\EE_{q_{\infty}}[T(X)]
\; = \;
\EE_{q_{\infty, \gamma_\infty}}[T(X)]
\end{align}
where $q_{\infty, \gamma_{\infty}}(x) \propto q_{\infty}^{1-\gamma_{\infty}}(x) \, \pi^{\gamma_{\infty}}(x)$. In the identity above, $T: \bR^d \to \bR^{d+ d(d+1)/2}$ equals $T(x)=(x_i, x_i x_j)_{i \leq j}$, representing the sufficient statistic vector for a $d$-dimensional Gaussian distribution in its natural parametrization. Recall that the Gaussian family can be parametrized as $q_{\lambda}(x) = \exp(\bra{\lambda, T(x)}) / Z(\lambda)$ for natural parameter $\lambda \in \Lambda \subset \bR^{d+d(d+1)/2}$ and associated normalizing constant $Z(\lambda) > 0$. We remark that the following can be generalized to any natural exponential family. Condition~\eqref{eq.alpha.div} describes the stationary points of the $\alpha$-divergence functional $\lambda \mapsto \distance{\alpha}{\pi}{q_{\lambda}}$
\citep[Equation~(7)]{Hernandez-Lobato2016}.
Consequently, \eqref{eq.alpha.div} shows that the limiting Gaussian distribution $q_\infty$ is a stationary point of the $\alpha$-divergence functional $\lambda \mapsto \distance{\alpha}{\pi}{q_{\lambda}}$ when choosing $\alpha = \gamma_\infty$. Since $\distance{\alpha}{\pi}{q} \to \kl{\pi}{q}$ as $\alpha \to 1$ and $\distance{\alpha}{\pi}{q} \to \kl{\pi}{q}$ as $\alpha \to 0$, this result further indicates that large update parameters $\gamma_t$ are to be favored since minimizing $\kl{ \pi }{ q_t }$ is preferred over minimizing $\kl{ q_t }{ \pi }$.

\section{Applications}
\label{sec:examples}

This section compares the performance of DAIS with other approximations.
Additionally, Appendix~\ref{ap:inverse} considers an inverse problem where, without any problem-specific adjustments or reduced approximation accuracy, DAIS is faster than an approximation that exploits the structure of the problem.
In Algorithm~\ref{alg:DAIS},
We set the ESS threshold $N_{\ess}$ to $10^3$,
the importance sample size
$S$ to $10^5$ and the robustness parameter $c$ to $0.5$
unless otherwise specified.
We use the Python package JAX \citep{Bradbury2018} for automatic differentiation to obtain $\nabla\!_x \Phi_t(x)$ and for parallelization of importance samples across CPU cores.

\subsection{Two-dimensional Synthetic Examples}
\label{sec:2D}

\begin{figure}[tb]
\begin{center}
\includegraphics[height=0.35\textheight]{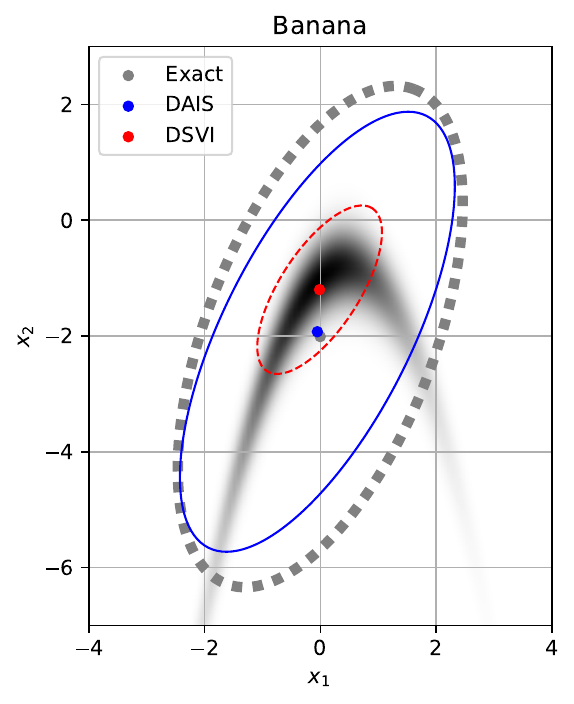}
\hspace{1cm}
\includegraphics[height=0.35\textheight]{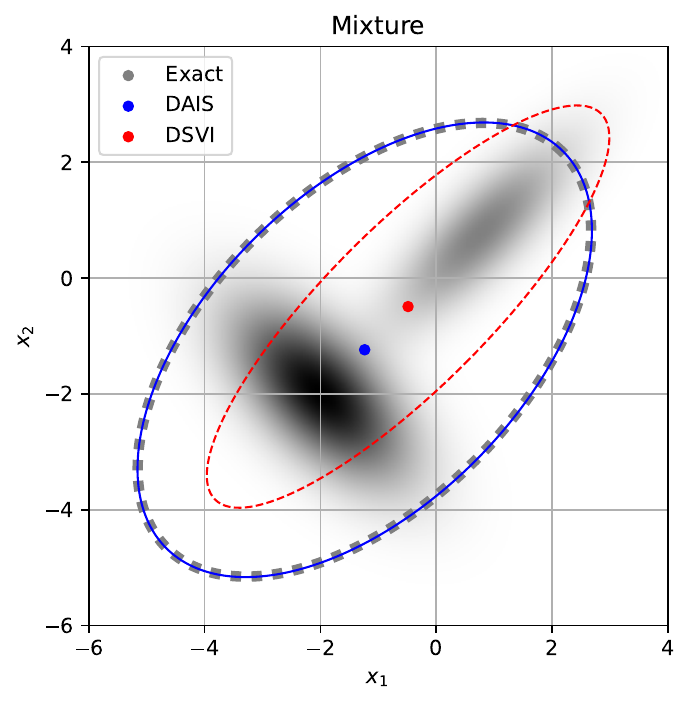}
\end{center}
\caption{The banana-shaped and mixture densities in grayscale with mean (dot) and 95\% credible regions (ellipse) of the corresponding Gaussian approximations overlaid.
The Gaussian approximations have their moments equal to the exact moments (dotted),
the DAIS estimates (solid) and the VI estimates (dashed).
\label{fig:banana_and_mixture}}
\end{figure}

As a first example, we consider two bivariate distributions from \citet{Ruiz2019} as their low dimensionality allows for easy inspection of approximations. Specifically, we consider the banana-shaped target distribution
\begin{equation}
    \pi(x) \propto \Normal\curBK{
    \begin{pmatrix}
        x_1 \\
        x_2 + x_1^2 + 1
    \end{pmatrix}
    \middle\vert\ 
    0, \begin{pmatrix}
        1 & 0.9 \\
        0.9 & 1
    \end{pmatrix}
    }
\end{equation}
and the mixture of two Gaussian distributions
\begin{equation}
    \pi(x) = 0.3\, \Normal\curBK{
    x\,
    \middle\vert\ 
    \begin{pmatrix}
        0.8 \\
        0.8
    \end{pmatrix}, \begin{pmatrix}
        1 & 0.8 \\
        0.8 & 1
    \end{pmatrix}
    }
    + 0.7\, \Normal\curBK{
    x\,
    \middle\vert\ 
    \begin{pmatrix}
        -2 \\
        -2
    \end{pmatrix}, \begin{pmatrix}
        1 & -0.6 \\
        -0.6 & 1
    \end{pmatrix}
    }
\end{equation}
visualized in Figure~\ref{fig:banana_and_mixture}.

We approximate the distributions using Gaussian proposals.
Algorithm~\ref{alg:DAIS}, i.e.\ DAIS, reaches $\gamma_t=1$ in 2 and 3 iterations for the banana-shaped and the mixture distribution, respectively.
Appendix~\ref{ap:2D_eps} considers a lower sample size $S$ resulting in a final $\gamma_t$ less than one. For comparison, we run VI with the same full covariance Gaussian distribution as DAIS uses.
Specifically, we minimize the reverse KL divergence $\kl{q_t}{\pi}$ via gradient descent.
Additionally, Appendix~\ref{ap:compare_2D} compares with adaptive IS methods.

Figure~\ref{fig:banana_and_mixture} summarizes the results. DAIS captures the mean and covariance of the target distribution $\pi$ more accurately than VI. In particular, the DAIS approximation for the mixture distribution is virtually indistinguishable from the exact mean and covariance. This shows the benefit of minimizing $\kl{\pi}{q_t}$ instead of $\kl{q_t}{\pi}$. The covariance underestimation by DAIS for the banana-shaped distribution is likely due to the ESS estimator in Step~\ref{step:S_eff} of Algorithm~\ref{alg:DAIS} underestimating MC error \citep{Martino2017,Elvira2022}.
Additionally, the results in Appendix~\ref{ap:2D_eps} where $\gamma_t < 1$ are in line with the fact that the adaptation of $\gamma_t$ interpolates between moment matching and VI.

\subsection{Logistic Regression}
\label{sec:logistic}

Lastly, we apply DAIS to the four logistic regression examples from Section~4.2.1 of \citet{Ong2018}.
Each data set consists of a binary response $y_i\in\{-1, 1\}$ and a $d$-dimensional feature vector $a_i\in\bR^d$ for $1\leq i\leq n$ where $n$ is the number of cases.
Then, the likelihood is $\ell(y\!\mid\! x) = \prod_{i=1}^n 1/\{1 + \exp(-y_i \bra{a_i, x})\}$
where $x$ is the coefficient vector. The prior on $x$ is $p_0(x) = \Normal(x\!\mid\!0, 10\,\mathrm{I}_d)$
such that the posterior follows as
$\pi(x)\propto \Normal\left(x\!\mid\! 0, 10\,\mathrm{I}_d\right) \ell(y\!\mid\! x)$.

The data involved are binarized versions of the spam, krkp, ionosphere and mushroom data from the UCI Machine Learning Repository \citep{Dua2019}. The binarization follows \citet[Section~5.1]{Gelman2008}. First, any continuous attributes are discretized using the method from \citet{Fayyad1993}.
Then, the resulting set of categorical attributes are encoded using dummy variables with the most frequent category as baseline. These dummy variables plus an intercept constitute the $d$ predictors considered. The resulting problem dimensionalities are summarized in Table~\ref{tab:uci}.

\begin{table}
\caption{Number of cases, number of categorical and continuous attributes, resulting number of predictors $d$, number of iterations and computation time in seconds of DAIS for the logistic regression examples.
\label{tab:uci}}
\begin{center}
\begin{tabular}{r|rrrr|rrr}
Data name & Cases & Categorical
& Continuous & $d$ & Iterations & Time \\
\hline
Spam & 4,601 & 57 & 0 & 105 & 7 & 24.3s \\
Krkp & 3,196 & 0 & 36 & 38 & 6 & 14.0s \\
Ionosphere & 351 & 32 & 0 & 111 & 12 & 11.3s \\
Mushroom & 8,124 & 0 & 22 & 96 & 10 & 58.5s \\
\end{tabular}
\end{center}
\end{table}

Algorithm~\ref{alg:DAIS}
approximates the mean and covariance of $\pi$
with
as initial approximation $q_1(x)=\Normal(x\!\mid\!\mu_1, \Gamma_1)$
the posterior mode $\mu_1 = \arg\max_x \pi(x)$ and the inverse Hessian of the negative log-density
$\Gamma_1 = \{-\nabla^2 U(\mu_1)\}^{-1}$.
Table~\ref{tab:uci} also lists computation times,
using an Intel i5-10600 CPU with six cores,
and number of DAIS iterations.
To assess approximation accuracy, we run Hamiltonian MC using the Python package 
BlackJAX \citep{cabezas2024blackjax} for 100,000 iterations, of which 10,000 are burn-in iterations.

\begin{figure}
\begin{center}
\includegraphics[width=\textwidth]{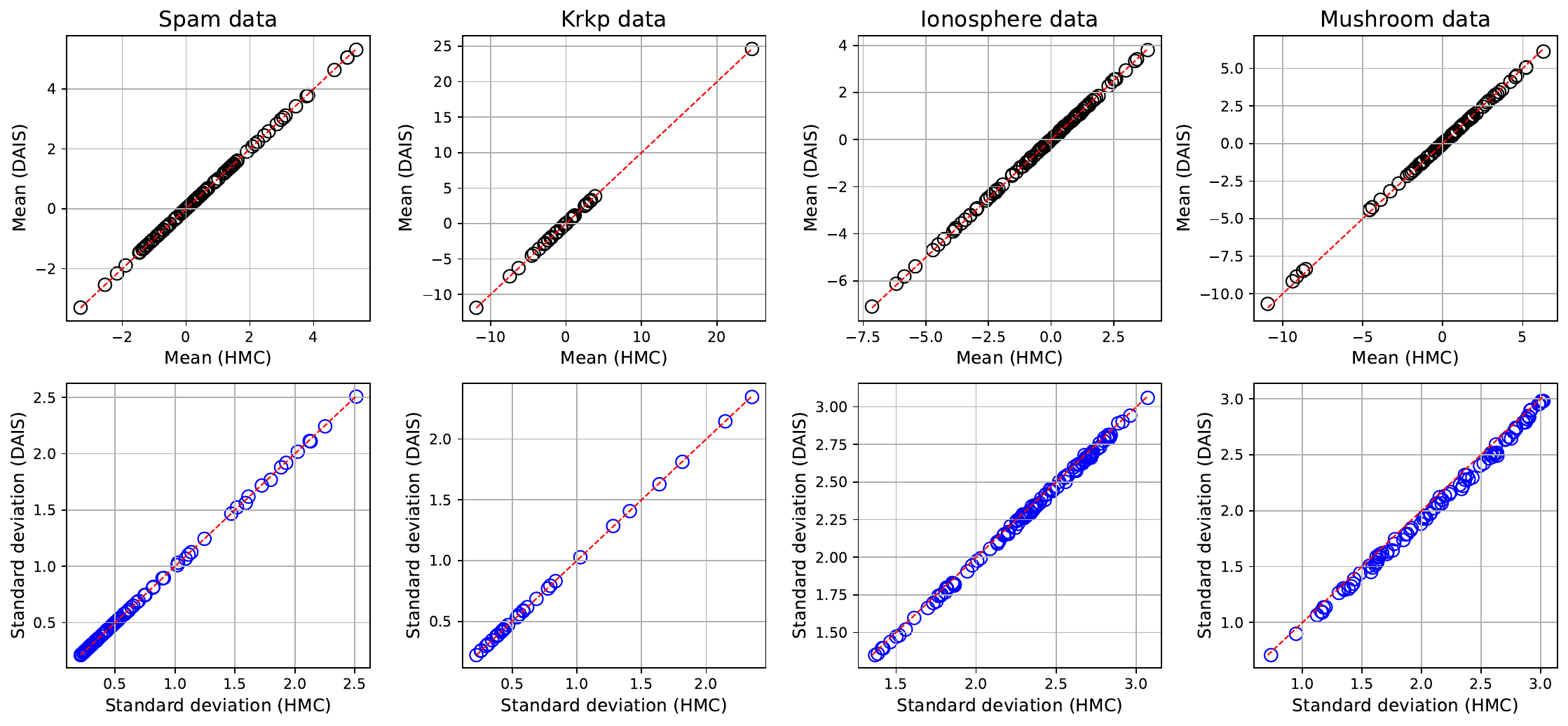}
\end{center}
\caption{
Scatter plots of the DAIS estimates versus the Hamiltonian MC estimates of the posterior means and standard deviations for the logistic regression examples.
\label{fig:uci}}
\end{figure}

Figure~\ref{fig:uci} shows that DAIS provides highly accurate estimates of posterior moments.
In Appendix~\ref{ap:compare_logistic}, the estimates are compared with variational inference with mean-field and full-covariance structure, as well as with adaptive multiple importance sampling \citep[AMIS,][]{Cornuet2012}.
To explore the effect of reduced MC error from
the application of Stein's identity (Section~\ref{sec.CV})
on the approximation of $\pi$,
Figure~\ref{fig:uci_no_Stein} in Supplementary Material
is the same as Figure~\ref{fig:uci}
except that it uses standard moment matching instead of the gradient-based updates in \eqref{eq.CV.mean.cov}.
Comparing these figures reveals that the methodology from Section~\ref{sec.CV} indeed improves approximation accuracy.

\section{Discussion}
\label{sec:discussion}

We propose a novel iterative approach to the approximation of the posterior distribution in general Bayesian models. The methodology is based on producing a sequence of Gaussian distributions whose moments match those of a damped target distribution, thus adapting to the target. The sequence is identified by exploiting Stein's lemma, which provides an updating rule for two consecutive sets of moments. The moments are computed via importance sampling while damping of the target is used to control the effective sample size (ESS) of the samples in the importance sampling.
The adaptation guarantees that the ESS is above a pre-specified threshold, which controls MC error, and provides a trade-off between minimizing the reverse and forward KL divergences based on computational constraints.
We call the method \emph{doubly adaptive importance sampling} (DAIS).
DAIS is a general methodology and competitive with methods that are more tailored to a problem-specific posterior. 

DAIS inherits certain limitations from IS. Firstly, high dimensionality typically results in too low ESS for IS to be feasible and translates to exceedingly high damping in DAIS.
Also, a large number of samples $S$ requires considerable computer memory.
At the same time, the standard estimate for ESS used here can underestimate MC error as alluded to in Section~\ref{sec:2D}, especially if the proposal $q_t$ is far from the target $\pi$.
See for instance \citet{Martino2017}, who also explore alternative measures of ESS
with markedly different behavior
such as the inverse of the maximum normalized IS weight: they find it to result in lower MC error at the same number of resampling steps in an SMC context.
Such alternative ESS measures are readily incorporated in DAIS.
Similarly,
schemes that minimize the MC error resulting from the self-normalization in IS could be explored \citep{Branchini2024}:
we have focused on self-normalized IS for simplicity.
Furthermore,
methods
for regularizing IS weights
can be applied
to smoothen the weights beyond what is achieved by damping
\citep{Vehtari2021}.

In the event that the IS estimate for the covariance $\Gamma_{t+1}$ in Algorithm~\ref{alg:DAIS} is not positive-definite, standard post-processing methods can be used for transforming the estimate into a positive-definite version of it. Possible approaches include setting the negative eigenvalues to small positive numbers. A more principled approach consists in reducing the damping parameter $\gamma_t$. Since the computational bottleneck generally lies in the evaluation of the target density $\pi$, recomputing the mean and covariance estimates for a reduced damping parameter $\gamma_t$ is generally computationally straightforward since no additional evaluation of the target density $\pi$ is necessary.

The fact that $q_t$
is Gaussian limits how well it
can approximate the target $\pi$.
To go beyond this limitation, $\pi$ can instead be approximated by the importance\hyp{}weighted samples from the last iteration of DAIS. The approximation can be made arbitrarily accurate by increasing the number of samples $S$ in this last iteration. Additionally, samples from multiple iterations of DAIS can be combined to approximate $\pi$ as in Equation~(17) of \citet{Bugallo2017},
especially if the amount of damping $\gamma_t$ is nearly constant over these iterations,
or by treating the proposals across iterations as components of a mixture proposal distribution as in \citet{Cornuet2012}.

Another avenue for increasing accuracy is going beyond Gaussianity for $q_t$. The Gaussian constraint
enables the MC error reduction in \eqref{eq.CV.mean.cov}. Other aspects of DAIS such as its adaptation
and the analysis in Section~\ref{sec:theory} do not require Gaussianity.
Moreover, \citet{Lin2019} extend Stein's lemma beyond Gaussian distributions to mixtures of an exponential family, potentially enabling MC error reduction similar to \eqref{eq.CV.mean.cov} for more general $q_t$. As such, ideas behind DAIS can be used with non-Gaussian approximating distributions.

In IS, it is typically desirable that the proposal has heavy tails to improve robustness. Therefore, one might consider a distribution $\overline{q}_{t} \ne q_t$
as proposal such as the
multivariate $t$-distribution
with location $\mu_t$ and scale $\Gamma_t$.
However, we have found that the choice can result in exceedingly small ESS for a high dimensionality $d$, and thus large amounts of damping.
After all, if $\pi$ is a posterior, it will often have Gaussian-like tails, e.g.\ because of Bernstein-von Mises.
Similarly, $\pi$ is then also likely unimodal.
Thus, we do not consider the multivariate $t$ or mixture distributions further.

\bigskip
\begin{center}
{\large\bf SUPPLEMENTARY MATERIAL}
\end{center}

\begin{description}
\item[Supplement:] Derivations,
discussion on monitoring convergence of Algorithm~\ref{alg:DAIS} and additional empirical results. (DAIS\_supplement.pdf, PDF file)
\item[Code:] Scripts that produce the empirical results along with a readme file are available at
\if0\blind
\url{https://github.com/thiery-lab/dais}. (DAIS\_code.zip, GitHub repository)
\fi
\if1\blind
[blinded]. (GitHub repository)
\fi
\end{description}

\noindent
The authors report there are no competing interests to declare.

\bibliographystyle{chicago_no_eds_no_month}  
\bibliography{DAIS}

\begin{thebibliography}{}

\bibitem[\protect\citeauthoryear{Agapiou, Papaspiliopoulos, Sanz-Alonso, and
  Stuart}{Agapiou et~al.}{2017}]{Agapiou2017}
Agapiou, S., O.~Papaspiliopoulos, D.~Sanz-Alonso, and A.~M. Stuart (2017).
\newblock Importance sampling: Intrinsic dimension and computational cost.
\newblock {\em Statistical Science\/}~{\em 32\/}(3), 405--431.

\bibitem[\protect\citeauthoryear{Akyildiz and Míguez}{Akyildiz and
  Míguez}{2021}]{Akyildiz2021}
Akyildiz, O.~D. and J.~Míguez (2021).
\newblock Convergence rates for optimised adaptive importance samplers.
\newblock {\em Statistics and Computing\/}~{\em 31\/}(2), 12.

\bibitem[\protect\citeauthoryear{Amari}{Amari}{1998}]{amari1998natural}
Amari, S. (1998).
\newblock Natural gradient works efficiently in learning.
\newblock {\em Neural Computation\/}~{\em 10\/}(2), 251--276.

\bibitem[\protect\citeauthoryear{Beskos, Jasra, Kantas, and Thiery}{Beskos
  et~al.}{2016}]{Beskos2016}
Beskos, A., A.~Jasra, N.~Kantas, and A.~Thiery (2016).
\newblock On the convergence of adaptive sequential {M}onte {C}arlo methods.
\newblock {\em The Annals of Applied Probability\/}~{\em 26\/}(2), 1111--1146.

\bibitem[\protect\citeauthoryear{Bishop}{Bishop}{2006}]{Bishop2006}
Bishop, C.~M. (2006).
\newblock {\em Pattern Recognition and Machine Learning}.
\newblock Information Science and Statistics. Springer, New York.

\bibitem[\protect\citeauthoryear{Blei, Kucukelbir, and McAuliffe}{Blei
  et~al.}{2017}]{Blei2017}
Blei, D.~M., A.~Kucukelbir, and J.~D. McAuliffe (2017).
\newblock Variational inference: A review for statisticians.
\newblock {\em Journal of the American Statistical Association\/}~{\em
  112\/}(518), 859--877.

\bibitem[\protect\citeauthoryear{Bradbury, Frostig, Hawkins, Johnson, Leary,
  Maclaurin, and Wanderman-Milne}{Bradbury et~al.}{2018}]{Bradbury2018}
Bradbury, J., R.~Frostig, P.~Hawkins, M.~J. Johnson, C.~Leary, D.~Maclaurin,
  and S.~Wanderman-Milne (2018).
\newblock {JAX}: composable transformations of {P}ython+{N}um{P}y programs.
\newblock \url{http://github.com/google/jax}.

\bibitem[\protect\citeauthoryear{Branchini and Elvira}{Branchini and
  Elvira}{2024}]{Branchini2024}
Branchini, N. and V.~Elvira (2024).
\newblock Generalizing self-normalized importance sampling with couplings.
\newblock arXiv:2406.19974v1.

\bibitem[\protect\citeauthoryear{Bugallo, Elvira, Martino, Luengo, Miguez, and
  Djuric}{Bugallo et~al.}{2017}]{Bugallo2017}
Bugallo, M.~F., V.~Elvira, L.~Martino, D.~Luengo, J.~Miguez, and P.~M. Djuric
  (2017).
\newblock Adaptive importance sampling: The past, the present, and the future.
\newblock {\em {IEEE} Signal Processing Magazine\/}~{\em 34\/}(4), 60--79.

\bibitem[\protect\citeauthoryear{Cabezas, Corenflos, Lao, Louf, Carnec,
  Chaudhari, et~al.}{Cabezas et~al.}{2024}]{cabezas2024blackjax}
Cabezas, A., A.~Corenflos, J.~Lao, R.~Louf, A.~Carnec, K.~Chaudhari, et~al.
  (2024).
\newblock {BlackJAX}: Composable {B}ayesian inference in {JAX}.
\newblock arXiv:2402.10797v2.

\bibitem[\protect\citeauthoryear{Chen, Huang, Huang, Reich, and Stuart}{Chen
  et~al.}{2023}]{Chen2024}
Chen, Y., D.~Z. Huang, J.~Huang, S.~Reich, and A.~M. Stuart (2023).
\newblock Gradient flows for sampling: Mean-field models, {G}aussian
  approximations and affine invariance.
\newblock arXiv:2302.11024v7.

\bibitem[\protect\citeauthoryear{Chopin, Crucinio, and Korba}{Chopin
  et~al.}{2024}]{Chopin2024}
Chopin, N., F.~R. Crucinio, and A.~Korba (2024).
\newblock A connection between tempering and entropic mirror descent.
\newblock arXiv:2310.11914v3.

\bibitem[\protect\citeauthoryear{Chopin and Papaspiliopoulos}{Chopin and
  Papaspiliopoulos}{2020}]{Chopin2020}
Chopin, N. and O.~Papaspiliopoulos (2020).
\newblock {\em An Introduction to Sequential Monte Carlo}.
\newblock Springer Series in Statistics. Springer Nature Switzerland.

\bibitem[\protect\citeauthoryear{Cornuet, Marin, Mira, and Robert}{Cornuet
  et~al.}{2012}]{Cornuet2012}
Cornuet, J.-M., J.-M. Marin, A.~Mira, and C.~P. Robert (2012).
\newblock Adaptive multiple importance sampling.
\newblock {\em Scandinavian Journal of Statistics\/}~{\em 39\/}(4), 798–812.

\bibitem[\protect\citeauthoryear{Daudel, Douc, and Roueff}{Daudel
  et~al.}{2023}]{Daudel2023}
Daudel, K., R.~Douc, and F.~Roueff (2023).
\newblock Monotonic alpha-divergence minimisation for variational inference.
\newblock {\em Journal of Machine Learning Research\/}~{\em 24}, 62.

\bibitem[\protect\citeauthoryear{Dehaene}{Dehaene}{2016}]{Dehaene2016}
Dehaene, G.~P. (2016).
\newblock Expectation propagation performs a smoothed gradient descent.
\newblock arXiv:1612.05053v1.

\bibitem[\protect\citeauthoryear{{Del Moral}, Doucet, and Jasra}{{Del Moral}
  et~al.}{2006}]{DelMoral2006}
{Del Moral}, P., A.~Doucet, and A.~Jasra (2006).
\newblock Sequential {M}onte {C}arlo samplers.
\newblock {\em Journal of the Royal Statistical Society: Series B (Statistical
  Methodology)\/}~{\em 68\/}(3), 411--436.

\bibitem[\protect\citeauthoryear{Domke and Sheldon}{Domke and
  Sheldon}{2018}]{Domke2018}
Domke, J. and D.~R. Sheldon (2018).
\newblock Importance weighting and variational inference.
\newblock In {\em Advances in Neural Information Processing Systems},
  Volume~31. Curran Associates, Inc.

\bibitem[\protect\citeauthoryear{Dua and Graff}{Dua and Graff}{2017}]{Dua2019}
Dua, D. and C.~Graff (2017).
\newblock {UCI} {M}achine {L}earning {R}epository.
\newblock University of California, Irvine, School of Information and Computer
  Sciences, \url{http://archive.ics.uci.edu/ml}.

\bibitem[\protect\citeauthoryear{Elvira and Chouzenoux}{Elvira and
  Chouzenoux}{2022}]{Elvira2022b}
Elvira, V. and E.~Chouzenoux (2022).
\newblock Optimized population {M}onte {C}arlo.
\newblock {\em IEEE Transactions on Signal Processing\/}~{\em 70}, 2489–2501.

\bibitem[\protect\citeauthoryear{Elvira, Chouzenoux, Akyildiz, and
  Martino}{Elvira et~al.}{2023}]{Elvira2023}
Elvira, V., E.~Chouzenoux, O.~D. Akyildiz, and L.~Martino (2023).
\newblock Gradient-based adaptive importance samplers.
\newblock {\em Journal of the Franklin Institute\/}~{\em 360\/}(13),
  9490–9514.

\bibitem[\protect\citeauthoryear{Elvira, Martino, and Robert}{Elvira
  et~al.}{2022}]{Elvira2022}
Elvira, V., L.~Martino, and C.~P. Robert (2022).
\newblock Rethinking the effective sample size.
\newblock {\em International Statistical Review\/}~{\em 90\/}(3), 525--550.

\bibitem[\protect\citeauthoryear{Fayyad and Irani}{Fayyad and
  Irani}{1993}]{Fayyad1993}
Fayyad, U.~M. and K.~B. Irani (1993).
\newblock Multi-interval discretization of continuous-valued attributes for
  classification learning.
\newblock In {\em Proceedings of the Thirteenth International Joint Conference
  on Artificial Intelligence, Vol. 2}, pp.\  1022--1027.

\bibitem[\protect\citeauthoryear{Gelman, Jakulin, Pittau, and Su}{Gelman
  et~al.}{2008}]{Gelman2008}
Gelman, A., A.~Jakulin, M.~G. Pittau, and Y.~Su (2008).
\newblock A weakly informative default prior distribution for logistic and
  other regression models.
\newblock {\em The Annals of Applied Statistics\/}~{\em 2\/}(4), 1360--1383.

\bibitem[\protect\citeauthoryear{Griewank and Walther}{Griewank and
  Walther}{2008}]{Griewank2008}
Griewank, A. and A.~Walther (2008).
\newblock {\em Evaluating Derivatives: Principles and Techniques of Algorithmic
  Differentiation\/} (2nd ed.).
\newblock Society for Industrial and Applied Mathematics, Philadelphia, PA.

\bibitem[\protect\citeauthoryear{Grosse, Maddison, and Salakhutdinov}{Grosse
  et~al.}{2013}]{Grosse2013}
Grosse, R.~B., C.~J. Maddison, and R.~R. Salakhutdinov (2013).
\newblock Annealing between distributions by averaging moments.
\newblock In {\em Advances in Neural Information Processing Systems},
  Volume~26. Curran Associates, Inc.

\bibitem[\protect\citeauthoryear{Guilmeau, Branchini, Chouzenoux, and
  Elvira}{Guilmeau et~al.}{2024a}]{Guilmeau2024}
Guilmeau, T., N.~Branchini, E.~Chouzenoux, and V.~Elvira (2024a).
\newblock Adaptive importance sampling for heavy-tailed distributions via
  $\alpha$-divergence minimization.
\newblock In {\em Proceedings of The 27th International Conference on
  Artificial Intelligence and Statistics}, Volume 238 of {\em Proceedings of
  Machine Learning Research}, pp.\  3871--3879. PMLR.

\bibitem[\protect\citeauthoryear{Guilmeau, Chouzenoux, and Elvira}{Guilmeau
  et~al.}{2024b}]{Guilmeau2024b}
Guilmeau, T., E.~Chouzenoux, and V.~Elvira (2024b).
\newblock Regularized {R}ényi divergence minimization through {B}regman
  proximal gradient algorithms.
\newblock arXiv:2211.04776v4.

\bibitem[\protect\citeauthoryear{Han and Liu}{Han and Liu}{2017}]{Han2017}
Han, J. and Q.~Liu (2017).
\newblock Stein variational adaptive importance sampling.
\newblock In {\em Proceedings of the Thirty-Third Conference on Uncertainty in
  Artificial Intelligence, {UAI} 2017, Sydney, Australia, August 11-15, 2017}.
  {AUAI} Press.

\bibitem[\protect\citeauthoryear{Hernández-Lobato, Li, Rowland, Bui,
  Hernández-Lobato, and Turner}{Hernández-Lobato
  et~al.}{2016}]{Hernandez-Lobato2016}
Hernández-Lobato, J., Y.~Li, M.~Rowland, T.~Bui, D.~Hernández-Lobato, and
  R.~Turner (2016).
\newblock Black-box $\alpha$-divergence minimization.
\newblock In {\em Proceedings of The 33rd International Conference on Machine
  Learning}, Volume~48 of {\em Proceedings of Machine Learning Research}, New
  York, NY, pp.\  1511--1520. PMLR.

\bibitem[\protect\citeauthoryear{Jerfel, Wang, Wong-Fannjiang, Heller, Ma, and
  Jordan}{Jerfel et~al.}{2021}]{Jerfel2021}
Jerfel, G., S.~Wang, C.~Wong-Fannjiang, K.~A. Heller, Y.~Ma, and M.~I. Jordan
  (2021).
\newblock Variational refinement for importance sampling using the forward
  {K}ullback-{L}eibler divergence.
\newblock In {\em Proceedings of the Thirty-Seventh Conference on Uncertainty
  in Artificial Intelligence}, Volume 161 of {\em Proceedings of Machine
  Learning Research}, pp.\  1819--1829. PMLR.

\bibitem[\protect\citeauthoryear{Khan and Nielsen}{Khan and
  Nielsen}{2018}]{Khan2018}
Khan, M.~E. and D.~Nielsen (2018).
\newblock Fast yet simple natural-gradient descent for variational inference in
  complex models.
\newblock In {\em 2018 International Symposium on Information Theory and Its
  Applications ({ISITA})}, pp.\  31--35. {IEEE}.

\bibitem[\protect\citeauthoryear{Koblents and M{\'{\i}}guez}{Koblents and
  M{\'{\i}}guez}{2013}]{Koblents2013}
Koblents, E. and J.~M{\'{\i}}guez (2013).
\newblock A population {M}onte {C}arlo scheme with transformed weights and its
  application to stochastic kinetic models.
\newblock {\em Statistics and Computing\/}~{\em 25\/}(2), 407--425.

\bibitem[\protect\citeauthoryear{Li and Turner}{Li and Turner}{2016}]{Li2016}
Li, Y. and R.~E. Turner (2016).
\newblock R\'{e}nyi divergence variational inference.
\newblock In {\em Advances in Neural Information Processing Systems},
  Volume~29. Curran Associates, Inc.

\bibitem[\protect\citeauthoryear{Lin, Khan, and Schmidt}{Lin
  et~al.}{2019}]{Lin2019}
Lin, W., M.~E. Khan, and M.~Schmidt (2019).
\newblock Stein's lemma for the reparameterization trick with exponential
  family mixtures.
\newblock arXiv:1910.13398v1.

\bibitem[\protect\citeauthoryear{Liu and Wang}{Liu and Wang}{2016}]{Liu2016}
Liu, Q. and D.~Wang (2016).
\newblock {S}tein variational gradient descent: A general purpose {B}ayesian
  inference algorithm.
\newblock In {\em Advances in Neural Information Processing Systems 29}. Curran
  Associates, Inc.

\bibitem[\protect\citeauthoryear{Martens}{Martens}{2020}]{Martens20}
Martens, J. (2020).
\newblock New insights and perspectives on the natural gradient method.
\newblock {\em Journal of Machine Learning Research\/}~{\em 21}, 146.

\bibitem[\protect\citeauthoryear{Martino, Elvira, and Louzada}{Martino
  et~al.}{2017}]{Martino2017}
Martino, L., V.~Elvira, and F.~Louzada (2017).
\newblock Effective sample size for importance sampling based on discrepancy
  measures.
\newblock {\em Signal Processing\/}~{\em 131}, 386–401.

\bibitem[\protect\citeauthoryear{Minka}{Minka}{2004}]{Minka2004}
Minka, T. (2004).
\newblock Power {EP}.
\newblock Technical Report MSR-TR-2004-149, Microsoft Research Ltd., Cambridge,
  UK.

\bibitem[\protect\citeauthoryear{Minka}{Minka}{2005}]{Minka2005}
Minka, T. (2005).
\newblock Divergence measures and message passing.
\newblock Technical Report MSR-TR-2005-173, Microsoft Research Ltd., Cambridge,
  UK.

\bibitem[\protect\citeauthoryear{Minka}{Minka}{2001}]{Minka2001}
Minka, T.~P. (2001).
\newblock Expectation propagation for approximate {B}ayesian inference.
\newblock In {\em Proceedings of the Seventeenth Conference on Uncertainty in
  Artificial Intelligence}, pp.\  362--369.

\bibitem[\protect\citeauthoryear{Mira, Solgi, and Imparato}{Mira
  et~al.}{2013}]{mira2013zero}
Mira, A., R.~Solgi, and D.~Imparato (2013).
\newblock Zero variance {M}arkov chain {M}onte {C}arlo for {B}ayesian
  estimators.
\newblock {\em Statistics and Computing\/}~{\em 23\/}(5), 653--662.

\bibitem[\protect\citeauthoryear{Neal}{Neal}{2001}]{Neal2001}
Neal, R.~M. (2001).
\newblock Annealed importance sampling.
\newblock {\em Statistics and Computing\/}~{\em 11\/}(2), 125--139.

\bibitem[\protect\citeauthoryear{Oates, Girolami, and Chopin}{Oates
  et~al.}{2017}]{Oates2017}
Oates, C.~J., M.~Girolami, and N.~Chopin (2017).
\newblock Control functionals for {M}onte {C}arlo integration.
\newblock {\em Journal of the Royal Statistical Society: Series B (Statistical
  Methodology)\/}~{\em 79\/}(3), 695--718.

\bibitem[\protect\citeauthoryear{Ong, Nott, and Smith}{Ong
  et~al.}{2018}]{Ong2018}
Ong, V. M.-H., D.~J. Nott, and M.~S. Smith (2018).
\newblock Gaussian variational approximation with a factor covariance
  structure.
\newblock {\em Journal of Computational and Graphical Statistics\/}~{\em
  27\/}(3), 465--478.

\bibitem[\protect\citeauthoryear{Opper and Winther}{Opper and
  Winther}{2005}]{Opper2005}
Opper, M. and O.~Winther (2005).
\newblock Expectation consistent approximate inference.
\newblock {\em Journal of Machine Learning Research\/}~{\em 6}, 2177--2204.

\bibitem[\protect\citeauthoryear{Paananen, Piironen, B\"{u}rkner, and
  Vehtari}{Paananen et~al.}{2021}]{Paananen2021}
Paananen, T., J.~Piironen, P.-C. B\"{u}rkner, and A.~Vehtari (2021).
\newblock Implicitly adaptive importance sampling.
\newblock {\em Statistics and Computing\/}~{\em 31\/}(2), 16.

\bibitem[\protect\citeauthoryear{Prangle and Viscardi}{Prangle and
  Viscardi}{2022}]{Prangle2022}
Prangle, D. and C.~Viscardi (2022).
\newblock Distilling importance sampling.
\newblock arXiv:1910.03632v4.

\bibitem[\protect\citeauthoryear{Ruiz and Titsias}{Ruiz and
  Titsias}{2019}]{Ruiz2019}
Ruiz, F. and M.~Titsias (2019).
\newblock A contrastive divergence for combining variational inference and
  {MCMC}.
\newblock In {\em Proceedings of the 36th International Conference on Machine
  Learning}, Volume~97 of {\em Proceedings of Machine Learning Research}, pp.\
  5537--5545. PMLR.

\bibitem[\protect\citeauthoryear{Ryu and Boyd}{Ryu and Boyd}{2015}]{Ryu2015}
Ryu, E.~K. and S.~P. Boyd (2015).
\newblock Adaptive importance sampling via stochastic convex programming.
\newblock arXiv:1412.4845v2.

\bibitem[\protect\citeauthoryear{Sanz-Alonso}{Sanz-Alonso}{2018}]{SanzAlonso2018}
Sanz-Alonso, D. (2018).
\newblock Importance sampling and necessary sample size: An information theory
  approach.
\newblock {\em SIAM/ASA Journal on Uncertainty Quantification\/}~{\em 6\/}(2),
  867–879.

\bibitem[\protect\citeauthoryear{Schuster}{Schuster}{2015}]{Schuster2015}
Schuster, I. (2015).
\newblock Gradient importance sampling.

\bibitem[\protect\citeauthoryear{Stein}{Stein}{1972}]{stein1972bound}
Stein, C. (1972).
\newblock A bound for the error in the normal approximation to the distribution
  of a sum of dependent random variables.
\newblock In {\em Proceedings of the sixth Berkeley symposium on mathematical
  statistics and probability, volume 2: Probability theory}, Volume~6, pp.\
  583--603. University of California Press.

\bibitem[\protect\citeauthoryear{Stuart}{Stuart}{2010}]{Stuart2010}
Stuart, A.~M. (2010).
\newblock Inverse problems: A {B}ayesian perspective.
\newblock {\em Acta Numerica\/}~{\em 19}, 451--559.

\bibitem[\protect\citeauthoryear{Vehtari, Gelman, Sivula, Jyl{\"a}nki, Tran,
  Sahai, Blomstedt, Cunningham, Schiminovich, and Robert}{Vehtari
  et~al.}{2020}]{Vehtari2020}
Vehtari, A., A.~Gelman, T.~Sivula, P.~Jyl{\"a}nki, D.~Tran, S.~Sahai,
  P.~Blomstedt, J.~P. Cunningham, D.~Schiminovich, and C.~P. Robert (2020).
\newblock Expectation propagation as a way of life: A framework for {B}ayesian
  inference on partitioned data.
\newblock {\em Journal of Machine Learning Research\/}~{\em 21}, 17.

\bibitem[\protect\citeauthoryear{Vehtari, Simpson, Gelman, Yao, and
  Gabry}{Vehtari et~al.}{2024}]{Vehtari2021}
Vehtari, A., D.~Simpson, A.~Gelman, Y.~Yao, and J.~Gabry (2024).
\newblock Pareto smoothed importance sampling.
\newblock arXiv:1507.02646v9.

\bibitem[\protect\citeauthoryear{Wang, Liu, and Liu}{Wang
  et~al.}{2018}]{Wang2018}
Wang, D., H.~Liu, and Q.~Liu (2018).
\newblock Variational inference with tail-adaptive $f$-divergence.
\newblock In {\em Advances in Neural Information Processing Systems 31}. Curran
  Associates, Inc.

\bibitem[\protect\citeauthoryear{Wiegerinck and Heskes}{Wiegerinck and
  Heskes}{2003}]{Wiegerinck2003}
Wiegerinck, W. and T.~Heskes (2003).
\newblock Fractional belief propagation.
\newblock In {\em Advances in Neural Information Processing Systems 15}. MIT
  Press.

\bibitem[\protect\citeauthoryear{Yao, Vehtari, Simpson, and Gelman}{Yao
  et~al.}{2018}]{Yao2018}
Yao, Y., A.~Vehtari, D.~Simpson, and A.~Gelman (2018).
\newblock Yes, but did it work?: Evaluating variational inference.
\newblock In {\em Proceedings of the 35th International Conference on Machine
  Learning}, Volume~80 of {\em Proceedings of Machine Learning Research}, pp.\
  5581--5590. PMLR.

\end{thebibliography}

\end{document}


\title{\bf Supplement to\\ ``Doubly Adaptive Importance Sampling'' published in the Journal of Computational and Graphical Statistics}
\author{\if0\blind Willem van den Boom, Andrea Cremaschi and Alexandre H.\ Thiery\fi}
\date{}
\maketitle

\appendix
\renewcommand{\theequation}{S\arabic{equation}}
\renewcommand{\thefigure}{S\arabic{figure}}

\section{Derivation of Equation~\texorpdfstring{\eqref{eq.CV.mean.cov}}{(2)}}
\label{ap:Stein}

For the standard orthonormal basis $(e_1, \ldots, e_d)$ of $\bR^d$, consider the constant test functions $\phi^{[i]}(x) = e_i$ for $1 \leq i \leq d$. An application of Stein's identity \eqref{eq.stein} to these functions $\phi^{[i]}$
gives $\EE_{q_{t, \gamma_t}}[\nabla \log q_{t, \gamma_t}(X)] = 0$.
The first line of
Equation~\eqref{eq.CV.mean.cov} follows now from
\begin{equation} \label{eq:grad_log_q}
    \nabla \log q_{t, \gamma_t}(x) = -\Gamma_t^{-1}{(x-\mu_t)} + \gamma_t \nabla \Phi_t(x).
\end{equation}

Similar
manipulations of \eqref{eq.stein} show that for a function $F: \bR^d \to \bR^d$ with Jacobian matrix $\jac_F(x)_{ij} = \partial_{x_j} F_i(x)$ and using test functions $\phi^{[ij]}(x)=e_i F_j(x)$,
\begin{align} \label{eq.stein.matrix}
\EE_{q_{t, \gamma_t}} \sqBK{ \nabla \log q_{t, \gamma_t}(X)  \otimes F(X) 
\, + \jac^\top_F(X) } 
\; = \; 0_{d\times d}.
\end{align}
Inserting
\eqref{eq:grad_log_q}
and
$F(X) = X - \mu_{t,\gamma_t}$
where
$\mu_{t,\gamma_t} = \EE_{q_{t, \gamma_t}}[X]$
yields that
\[
    \EE_{q_{t, \gamma_t}}[(X-\mu_{t})\otimes (X - \mu_{t,\gamma_t})]
    = \Gamma_t + \gamma_t\Gamma_t\, \EE_{q_{t, \gamma_t}}[\nabla \Phi_t(X) \otimes (X - \mu_{t,\gamma_t})]
\]
from which the second line of \eqref{eq.CV.mean.cov} follows.

\section{Proof of Proposition~\ref{prop:rmse}}
\label{ap:rmse}

By Equation~(2.3) of \citet{Agapiou2017},
we have that self-normalized IS
estimation of
$\EE_{q_{t,\gamma_t}}[\phi(X)]$
with $S$ samples
from $q_t$
has an RMSE of $1/\sqrt{S}$ as 
$S\to\infty$
under the following two conditions:
\begin{align} \label{eq:conditions}
\EE_{q_t}[\{q_{t,\gamma_t}(X) / q_t(X)\}^2]
&\propto
\EE_{q_t}[e^{2\gamma_t \Phi_t(X)}]
<\infty \\
\EE_{q_t}[\{\phi(X)\, q_{t,\gamma_t}(X) / q_t(X)\}^2]
&\propto
\EE_{q_t}[\phi^2(X)\, e^{2\gamma_t \Phi_t(X)}]
<\infty
\end{align}
These conditions are satisfied in part~(ii) of Proposition~\ref{prop:rmse} by assumption.
Furthermore, the presence of the factor $\gamma_t$ in the right-hand side of \eqref{eq.CV.mean.cov} results in the asymptotic RMSE of order $\gamma_t/\sqrt{S}$.

For part~(i),
consider the unnormalized density
$f(x) = q_t(x)\, e^{2\gamma_t \Phi_t(x)}$.
Then,
\begin{multline}
    \EE_{q_t}[X_i^2\, e^{2\gamma_t \Phi_t(X)}]^2
=
\left\{\int x_i^2\, q_t(x)\, e^{2\gamma_t \Phi_t(x)} dx\right\}^2
= \EE_{f}[X_i^2]^2 \\
<
\EE_{f}[X_i^4]
= \int x_i^4\, q_t(x)\, e^{2\gamma_t \Phi_t(x)} dx
=
\EE_{q_t}[X_i^4\, e^{2\gamma_t \Phi_t(X)}]
\end{multline}
where the inequality follows from Jensen's inequality
and $\EE_{q_t}[X_i^4\, e^{2\gamma_t \Phi_t(X)}] < \infty$ by assumption.
Thus,
\eqref{eq:conditions} is also satisfied when considering self-normalized IS estimation of the left-hand side of the first line of \eqref{eq.CV.mean.cov}.
Furthermore, \eqref{eq:conditions} is satisfied for the left-hand side of the second line of \eqref{eq.CV.mean.cov} by assumption such that the required result follows.

\section{Monitoring Convergence}
\label{ap:monitoring}

In challenging settings where the target distribution departs significantly from Gaussianity, running DAIS with a fixed number of IS particles $S$ per iteration produces a sequence of damping parameters $\gamma_t$ that does not eventually converge to one. Furthermore, it is typically not feasible to reliably estimate the forward KL divergence $\kl{\pi}{q_t}$ with MC methods.
Although the trajectory $t \mapsto \gamma_t$ is typically noisy and not necessarily increasing, we observe that the damping parameter eventually (and often rapidly) reaches a stationary regime. For further monitoring of convergence, we track the Evidence Lower BOund
%
%
\begin{align} \label{eq.elbo}
\elbo(q_t) = \int_{x} \log\curBK{ \frac{\overline{\pi}(x)}{q_t(x)} } \, q_t(x) \, dx
\end{align}
where $\pi(x) = \overline{\pi}(x) / Z$ for an unknown normalization constant $Z>0$. Producing an estimate $\widehat{\elbo}$ of \eqref{eq.elbo} with importance sampling is straightforward since all quantities necessary for its evaluation would have typically already been evaluated while running the DAIS algorithm.

\begin{figure}[tb]
\begin{center}
\includegraphics[width=0.45\textwidth]{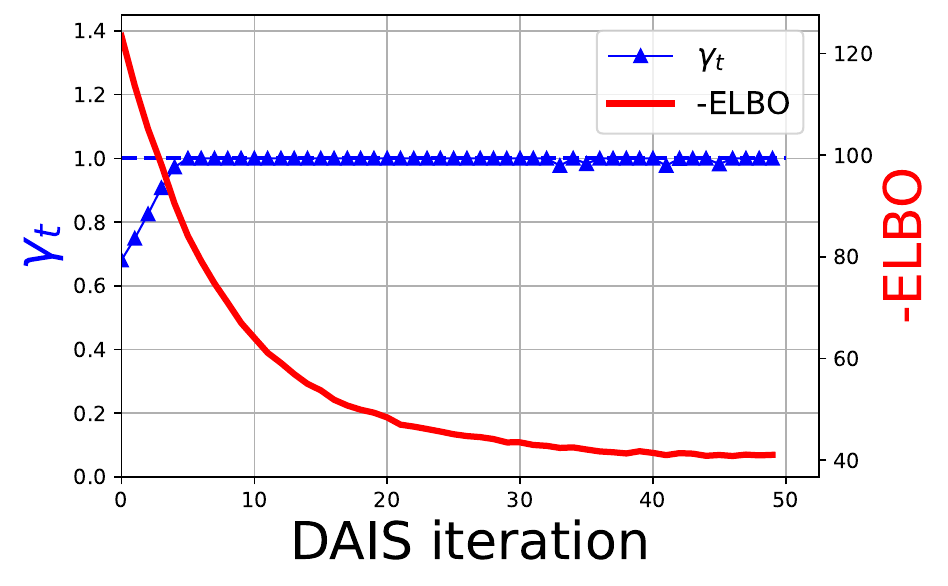}
\includegraphics[width=0.45\textwidth]{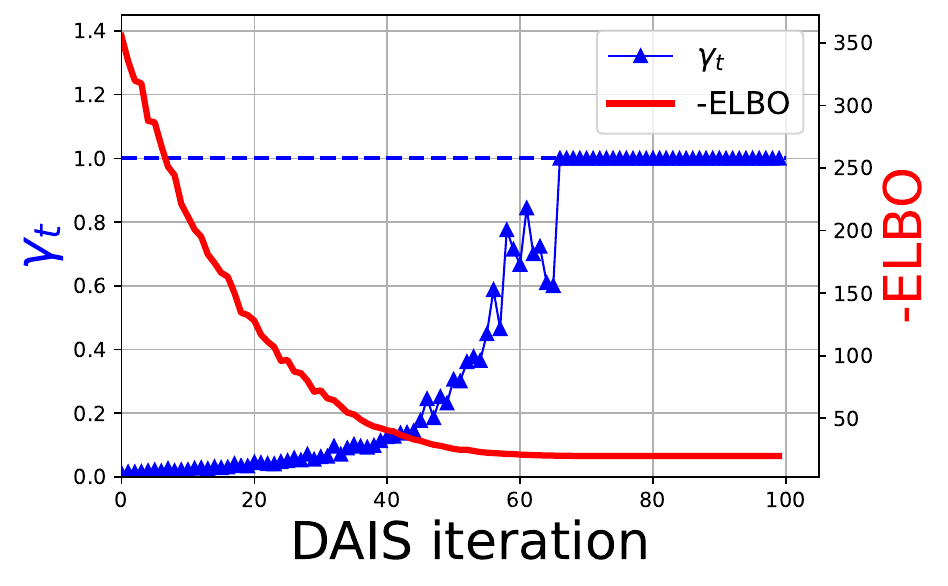}
\end{center}
\caption{Tracking of the damping parameter $\gamma_t$ and (estimate of) the negative $\elbo$ for monitoring convergence. {\bf (Left)} $2$-dimensional target $\pi(x) \propto {\exp[-\{x_2 - \mathcal{F}(x_1)\}^2 / (2 \sigma^2)]}\, p_0(x_1, x_2)$ with $\mathcal{F}(x) = 1 + \sin(2x)$ and $p_0(x_1, x_2)$ the density of a standard Gaussian density in $\mathbb{R}^2$. {\bf (Right)} $d=100$ dimensional Gaussian target distribution with mean $\mu=(1,\ldots,1) \in \bR^d$ and covariance $\Gamma_{i,j} = 0.9 + 0.1 \, \delta(i=j)$.}
\label{fig:monitor}
\end{figure}

Figure~\ref{fig:monitor} displays the trajectories of $t \mapsto \gamma_t$ and $t \mapsto -\widehat{\elbo}(q_t)$ when DAIS is used for approximating the following two target densities:
\begin{enumerate}[label=(\roman*), font=\upshape]
\item
a $d=2$ dimensional density $\pi(x_1, x_2)$ defined as
\[\frac{\pi(x_1, x_2)}{p_{0}(x_0,x_1)} \propto \exp\left[-\frac{\{x_2 - \mathcal{F}(x_1)\}^2}{ 2\sigma^2}\right]\] for $\sigma = 0.1$, non-linear function $\mathcal{F}(x) = 1 + \sin(2x)$ and standard Gaussian prior density $p_0(x_0, x_1)$;
\item
a $d=100$ dimensional Gaussian distribution with mean $\mu=(1,\ldots,1) \in \bR^d$ and covariance $\Gamma_{i,j} = 0.9 + 0.1 \, \delta(i=j)$.
\end{enumerate}
In both cases, DAIS is started from a standard multivariate Gaussian distribution, i.e.\ $\mu_1 = 0_{d\times 1}$ and $\Gamma_1 = \mathrm{I}_d$.
We use the robustness parameters of $c = 0.1$ and $c = 0.3$ for the 2-dimensional and the 100-dimensional target, respectively.
Figure~\ref{fig:monitor} shows that monitoring either the damping parameter $\gamma_t \in (0,1]$ or the ELBO leads to roughly the same conclusion.

\section{Derivations for Section~\texorpdfstring{\ref{sec:theory}}{4}}
\label{ap:theory}

A derivation of \eqref{eq.NG.descent} can be found in \citet{khan2017variational} and the equivalence between the two formulations follows from the chain rule. Since $\nabla_{\Gamma} \EE_{q}[\phi(X)] = \tfrac12 \, \EE_{q}[\nabla_{xx} \phi(X)]$ for any test function $\phi: \mathbb{R}^d \to \mathbb{R}$ \citep[Equation~(A.3)]{opper2009variational}, standard algebraic manipulations show that the forward (Fwd) and reverse (Rev) KL divergences satisfy:
\begin{align}
\textrm{(Rev)}:
&\left\{
\begin{aligned}
\nabla_\mu \kl{q}{\pi} &= -\EE_{q}[\nabla_{x} \log \pi(X)]\\
\nabla_\Gamma \kl{q}{\pi} &= \frac{1}{2}\left(-\EE_{q}[\nabla^2_{xx} \log \pi(X)] - \Gamma^{-1}\right)
\end{aligned}
\right.
\\
\textrm{(Fwd)}:
&\left\{
\begin{aligned}
\nabla_\mu \kl{\pi}{q} &= -\Gamma^{-1}\,\EE_{\pi}[(X-\mu)]\\
\nabla_\Gamma \kl{\pi}{q} &= -\frac{1}{2}\,\EE_{\pi}[\Gamma^{-1}(X-\mu)\otimes (X-\mu)\Gamma^{-1} - \Gamma^{-1}].
\end{aligned}
\right.
\end{align}
It follows that the natural\hyp{}gradient flow for minimizing the forward and reverse KL divergences are given by
\begin{align}
\textrm{(Rev)}:
\left\{
\begin{aligned}
\frac{d \mu}{dt} &= \Gamma\, \EE_{q}[\nabla_{x} \log \pi(X)]\\
\frac{d \Gamma}{dt} &= \Gamma\, \EE_{q}[\nabla^2_{xx} \log \pi(X)]\, \Gamma + \Gamma
\end{aligned}
\right.
\qquad
\textrm{(Fwd)}:
\left\{
\begin{aligned}
\frac{d \mu}{dt} &= \EE_{\pi}[(X-\mu)]\\
\frac{d \Gamma}{dt} &= \EE_{\pi}[(X-\mu)\otimes(X-\mu)] - \Gamma.
\end{aligned}
\right.
\end{align}
Since $q_{t, \gamma_t}$ converges to $q_t(x) = \Normal(x\!\mid\!\mu, \Gamma)$ as $\gamma_t \to 0$ and $\nabla \Phi_t(x) = \nabla \log \pi(x) + \Gamma^{-1} (x-\mu)$, the quantities $\Gg_\mu(q_{t, \gamma_t}) = \EE_{q_{t, \gamma_t}}[\Gamma \, \nabla \Phi_t(X)]$ and $\GG_\Gamma(q_{t, \gamma_t}) = \cov_{q_{t, \gamma_t}}[  \Gamma \, \nabla \Phi_t(X), X]$ satisfy \eqref{eq.NG.rev}:
the second equality in \eqref{eq.NG.rev} follows from an integration by parts (or Stein's lemma).
Furthermore, since $q_{t, \gamma_t}$ converges to $\pi$ as $\gamma_t \to 1$, the definitions $\mu_{t, \gamma_t} 
= \mu_t + \gamma_t \, \Gg_\mu(q_{t, \gamma_t})$ and $\Gamma_{t, \gamma_t} 
 = \Gamma_t + \gamma_t \, \GG_\Gamma(q_{t, \gamma_t})$ show \eqref{eq.NG.fwd}.

\cite{Amari1985} defines the $\alpha$-divergence as 
\begin{align}
\distance{\alpha}{\pi}{q_{\lambda}} = \frac{1}{\alpha (1-\alpha)} \sqBK{1 - \int_{\cX} \curBK{\frac{\pi(x)}{q_\lambda(x)} }^{\alpha} q_\lambda(x)\, dx }.
\end{align}
Since
$\nabla_\lambda q_\lambda(x) = q_\lambda(x)\{T(x) - \EE_{q_\lambda}[T(X)]\}$,
we obtain
\citep[Equation~(7)]{Hernandez-Lobato2016}
\begin{equation}
    \nabla_\lambda \distance{\alpha}{\pi}{q_{\lambda}} = \frac{Z(\lambda, \alpha)}{\alpha} \BK{ \EE_{q_\lambda}[T(X)] - \EE_{q_{\lambda, \alpha}}[T(X)] }
\end{equation}
where $Z(\lambda, \alpha) = \int \pi(x)^{\alpha}\, q^{1-\alpha}_{\lambda}(x)\, dx$.
Thus,
condition~\eqref{eq.alpha.div} describes the stationary points of the $\alpha$-divergence functional $\lambda \mapsto \distance{\alpha}{\pi}{q_{\lambda}}$.

\section{Two-dimensional Examples Converging to \texorpdfstring{$\gamma_t<1$}{ε < 1}}
\label{ap:2D_eps}

\begin{figure}
\begin{center}
\includegraphics[height=0.35\textheight]{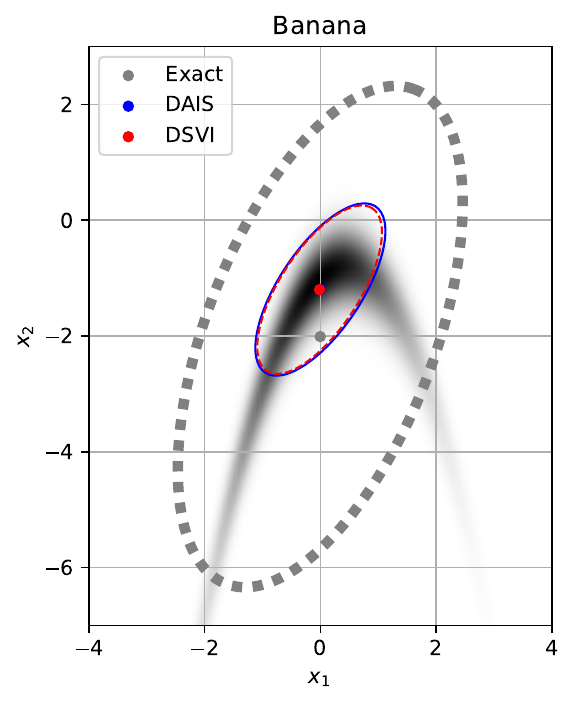}
\hspace{1cm}
\includegraphics[height=0.35\textheight]{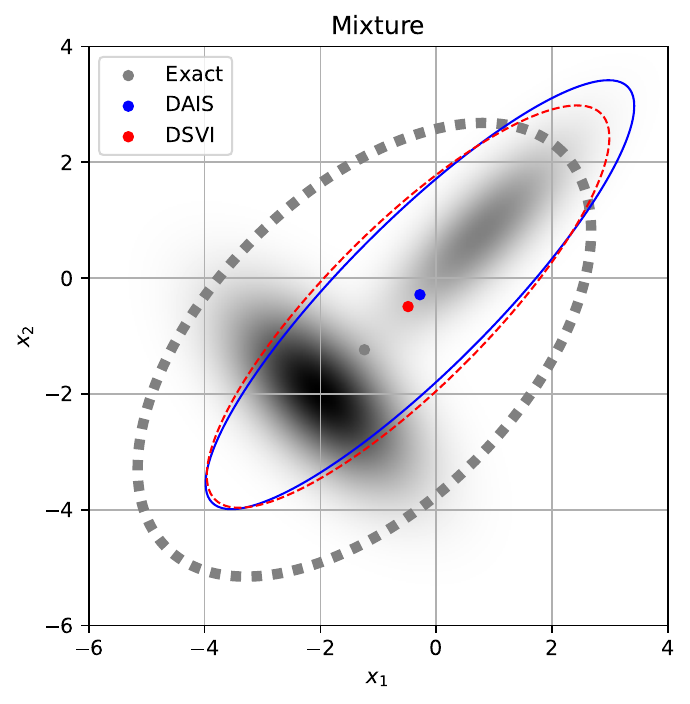}
\end{center}
\caption{The banana-shaped and mixture densities in grayscale with the mean (dot) and the 95\% credible regions (ellipse) of their Gaussian approximations overlaid.
The Gaussian approximations have their moments equal to the exact moments (gray, dotted),
the DAIS estimates using $S=1010$ resulting in $\gamma_t<1$ (blue, solid)
and the VI estimates (red, dashed).
\label{fig:banana_and_mixture_eps}}
\end{figure}

The set-up of Section~\ref{sec:2D}
uses importance sample size $S=10^5$ such that DAIS finishes with $\gamma_t=1$.
This appendix instead considers $S=1010$, which is only slightly higher than the effective sample size threshold $N_{\ess}=10^3$.
Then, DAIS
converges to
$\gamma_t\approx 0.14$
and
$\gamma_t\approx 0.10$
for the banana-shaped and mixture distributions, respectively.
Comparing Figure~\ref{fig:banana_and_mixture} in the main text with Figure~\ref{fig:banana_and_mixture_eps}
confirms that the adaptation of $\gamma_t$
indeed interpolates between moment matching and variational inference (VI).

\section{Comparisons to AMIS and OAIS: Two-dimensional Synthetic Examples}
\label{ap:compare_2D}

\begin{figure}
\begin{center}
\includegraphics[height=0.35\textheight]{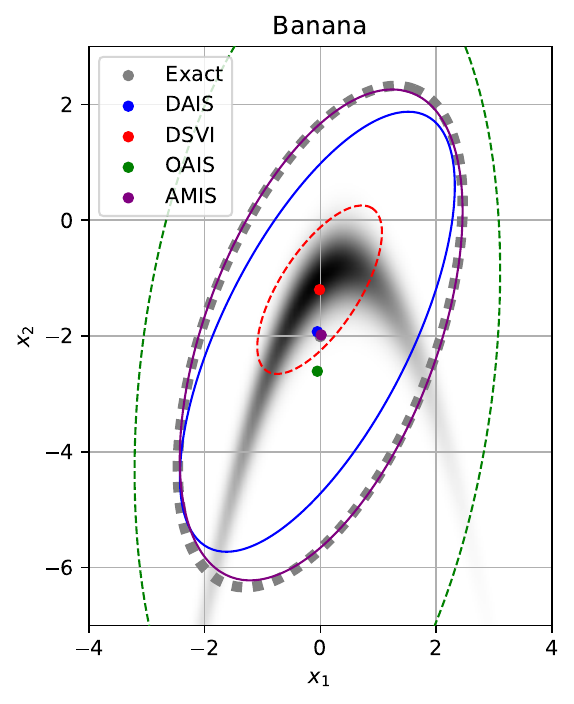}
\hspace{1cm}
\includegraphics[height=0.35\textheight]{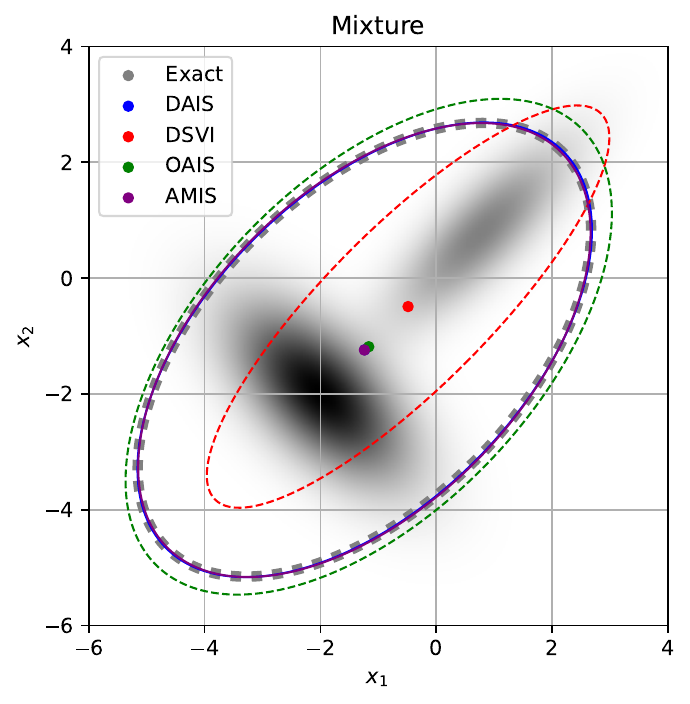}
\end{center}
\caption{The banana-shaped and mixture densities are shown in grayscale, with the mean (dot) and the 95\% credible regions (ellipse) of their Gaussian approximations overlaid. In these non-Gaussian examples, DSVI (red, dashed) fails to accurately approximate the second moments. In contrast, for the mixture density example, DAIS, OAIS and AMIS all provide good approximations of the second moments, with AMIS being indistinguishable from the ground truth. The severely non-Gaussian example of the banana-shaped density presents a greater challenge, with the methods approximating the second moments with varying degrees of accuracy.
\label{fig:banana_and_mixture_AMIS_OAIS}}
\end{figure}

In the two-dimensional examples presented in Section~\ref{sec:2D}, we compare DAIS with adaptive multiple importance sampling \citep[AMIS,][]{Cornuet2012} and optimized adaptive importance sampling \citep[OAIS,][]{Akyildiz2021}, as well as Gaussian variational inference fitted through the doubly stochastic variational inference (DSVI) approach of \citet{Titsias2014}. The resulting approximations are reported and discussed in Figure~\ref{fig:banana_and_mixture_AMIS_OAIS}.

AMIS is a foundational method in the field of multiple importance sampling and has inspired numerous subsequent works \citep{Bugallo2017}. For this comparison, AMIS is implemented using a Student's $t$-distribution with three degrees of freedom as the proposal family. At each of the $R = 20$ rounds of adaptation, $S = 10^5$ samples were generated.

The DAIS method is also compared to OAIS \citep{Akyildiz2021}, which optimizes the upper bound $R(\theta) = \mathbb{E}_{q_{\theta}}[\Pi^2(X)/q_{\theta}^2(X)]$ to minimize the mean square error (MSE) in self-normalized importance sampling. This approach employs a proposal density $q_{\theta}(x)$ to approximate expectations under the target density $\pi(x) = \Pi(x) / \mathcal{Z}$. While we successfully implemented OAIS for these low-dimensional examples, extending the method to higher-dimensional settings, such as the logistic regression examples considered in the next section, proved challenging. The primary difficulty lies in the ratio $\Pi^2(x)/q_{\theta}^2(x)$, which can span several orders of magnitude in high-dimensional scenarios, thereby complicating the adaptive procedure and its dependence on the normalization constant $\mathcal{Z}$. OAIS was implemented using $S=10^3$ samples and $10^4$ iterations of the ADAM optimizer with a learning rate of $10^{-2}$. We experimented with a range of learning rates and significantly increased the number of importance samples, but the results were relatively insensitive to these choices.

In contrast, we have found that VI methods, which optimize $\mathbb{E}_{q_{\theta}}[\log\{q_{\theta}(X) / \pi(X)\}]$, are both straightforward and stable to implement. Notably, the gradient $\nabla_{\theta} \mathbb{E}_{q_{\theta}}[\log\{q_{\theta}(X) / \pi(X)\}]$ does not depend on the normalizing constant $\mathcal{Z}$, and the logarithm stabilizes the optimization. Conversely, we have found it challenging to reliably minimize quantities such as $R(\theta) = \mathbb{E}_{q_{\theta}}[\Pi^2(X)/q_{\theta}^2(X)]$. While our proposed method also relies on quantities of the type $\Pi(x) / q(x)$, the adaptive damping procedure is indeed one key mechanism for stabilizing the adaptation procedure.

\section{Comparisons to AMIS and Variational Inference: Logistic Regression}
\label{ap:compare_logistic}

In the four logistic regression examples presented in Section~\ref{sec:logistic}, we compare DAIS with adaptive multiple importance sampling \citep[AMIS,][]{Cornuet2012}, as well as with variational inference using both a diagonal covariance matrix and a full covariance matrix. The VI methods were implemented using the doubly stochastic variational inference framework of \citet{Titsias2014}. As detailed in Section~\ref{ap:compare_2D}, we were unable to successfully apply the optimized adaptive importance sampling method of \citet{Akyildiz2021} in these high-dimensional settings. We report estimates of the marginal means and standard deviations, comparing them to the ground truth obtained from a Hamiltonian Monte Carlo (HMC) run at convergence.

\begin{figure}
\begin{center}
\includegraphics[width=\textwidth]{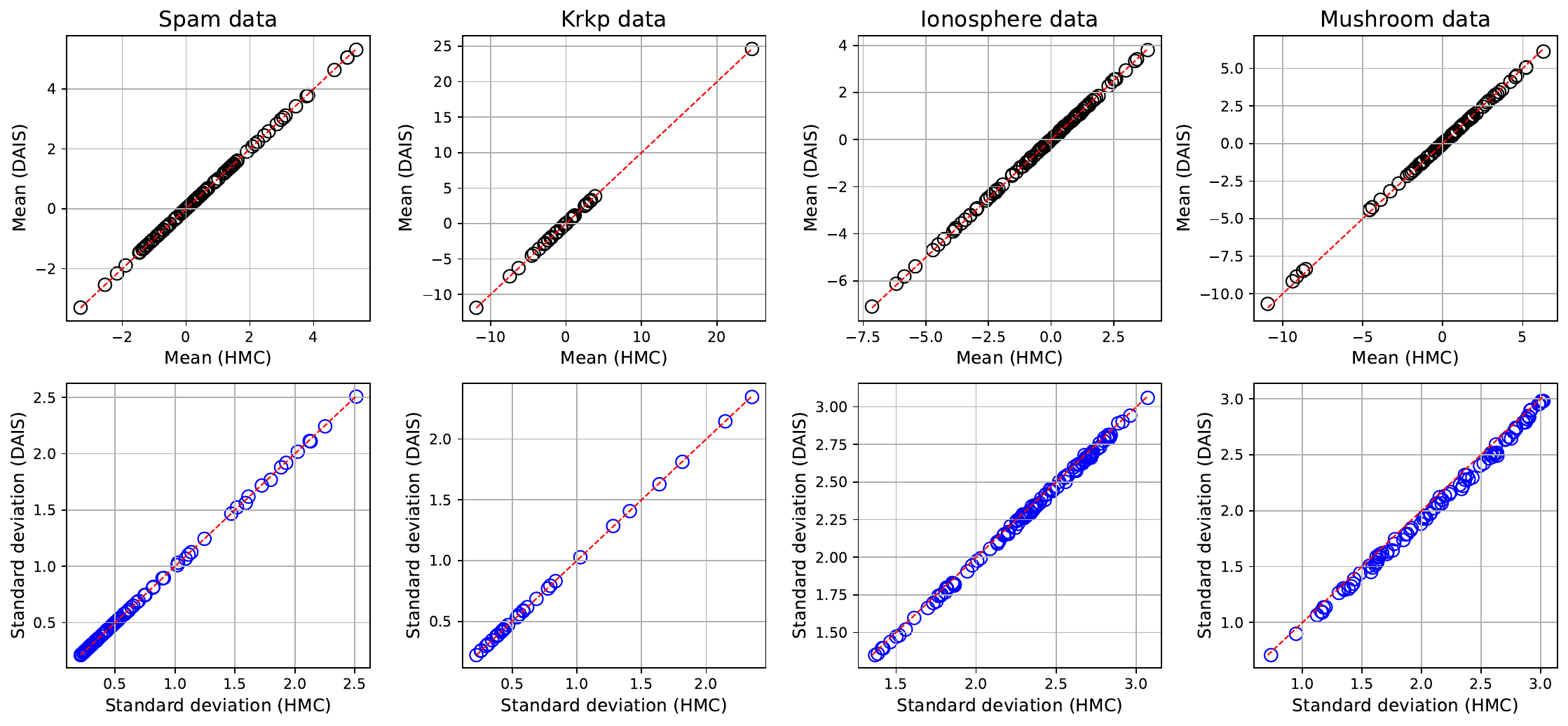}
\end{center}
\caption{
Scatter plots of the DAIS estimates versus the Hamiltonian MC estimates of the posterior means and standard deviations for the logistic regression examples.
\label{fig:uci_dais_vs_hmc}}
\end{figure}

\begin{figure}
\begin{center}
\includegraphics[width=\textwidth]{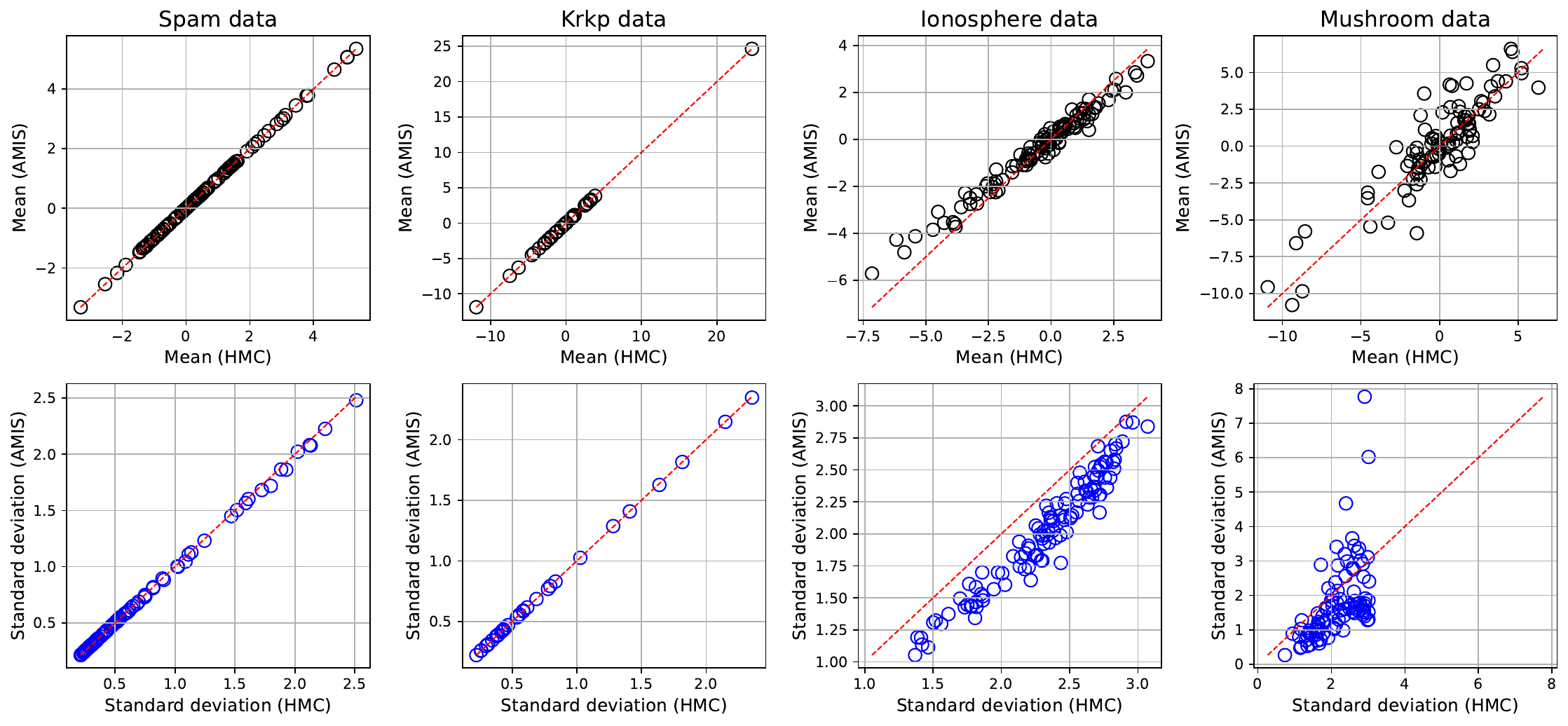}
\end{center}
\caption{
Scatter plots of the AMIS estimates versus the Hamiltonian MC estimates of the posterior means and standard deviations for the logistic regression examples.
\label{fig:uci_amis_vs_hmc}}
\end{figure}

\begin{figure}
\begin{center}
\includegraphics[width=\textwidth]{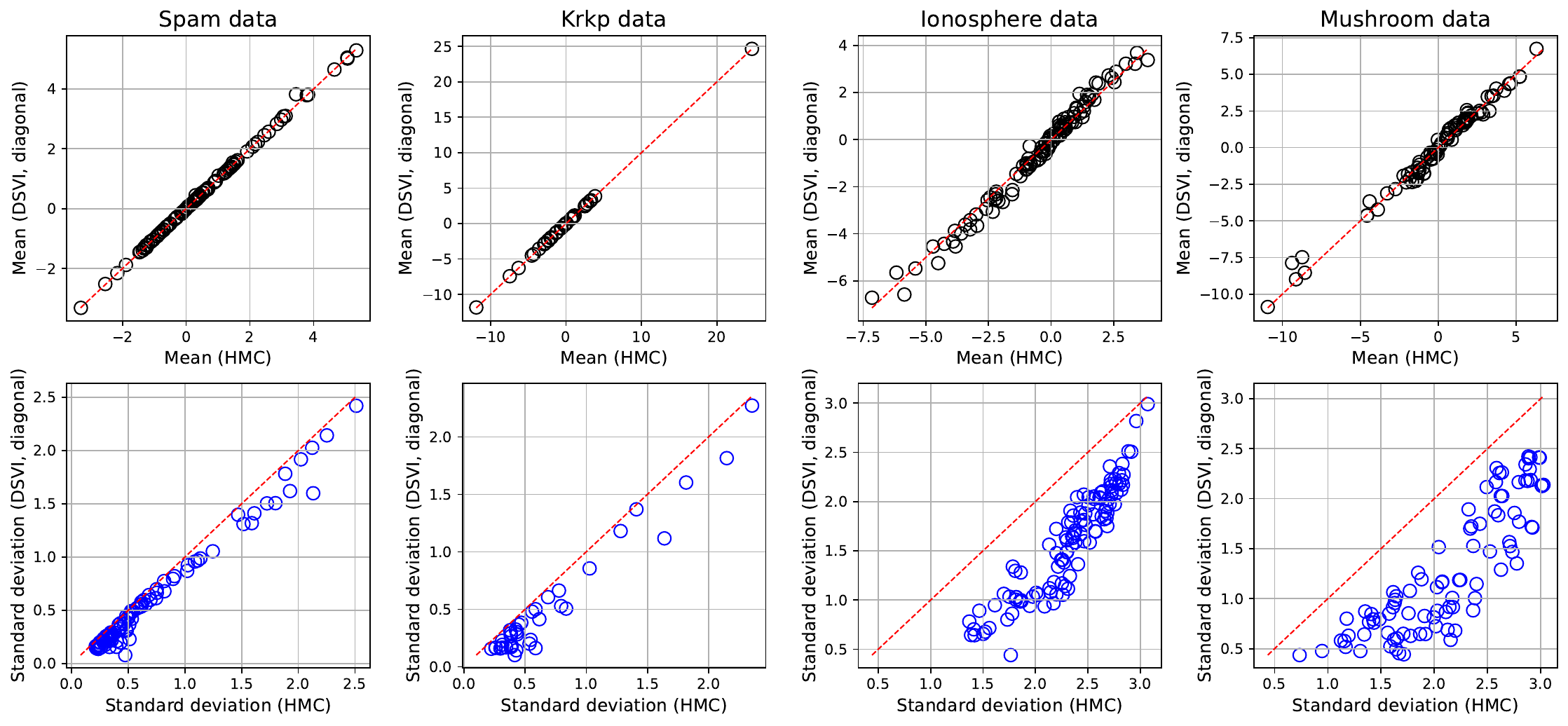}
\end{center}
\caption{
Scatter plots of the estimates from VI with diagonal covariance matrix versus the Hamiltonian MC estimates of the posterior means and standard deviations for the logistic regression examples.
\label{fig:uci_dsvi_diag_vs_hmc}}
\end{figure}

\begin{figure}
\begin{center}
\includegraphics[width=\textwidth]{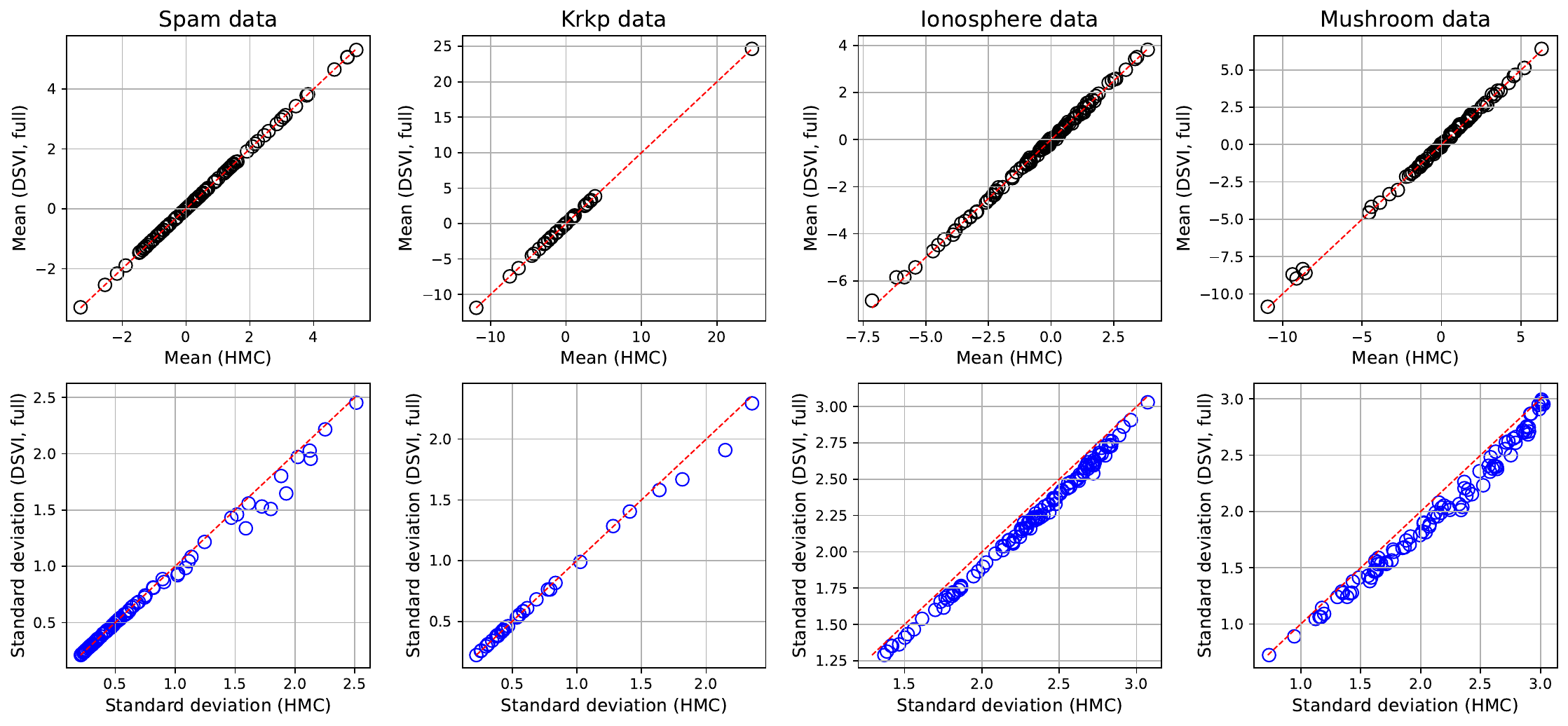}
\end{center}
\caption{
Scatter plots of the estimates from VI with full covariance matrix versus the Hamiltonian MC estimates of the posterior means and standard deviations for the logistic regression examples.
\label{fig:uci_dsvi_full_vs_hmc}}
\end{figure}

For DAIS (Figure~\ref{fig:uci_dais_vs_hmc}, same as Figure~\ref{fig:uci} in the main text), we used $S=10^5$ importance samples with a target effective sample size of $N_{\text{ESS}}=10^3$ and a robustness parameter of $c=0.5$
as mentioned in Section~\ref{sec:examples}. The AMIS method (Figure~\ref{fig:uci_amis_vs_hmc}) was implemented using a Student's $t$-distribution with three degrees of freedom as the proposal family. At each of the $R = 20$ adaptation rounds, $S = 10^5$ samples were generated. The VI methods (Figures~\ref{fig:uci_dsvi_diag_vs_hmc} and \ref{fig:uci_dsvi_full_vs_hmc}) were run for $10^3$ iterations, with $10^3$ Monte Carlo samples per iteration and an ADAM learning rate of $10^{-2}$. All methods were initialized using a Laplace approximation of the target distribution.

For the spam and krkp data, both DAIS and AMIS provide highly accurate estimates.
VI with full covariance is accurate as well though slightly less so. Finally, VI with a diagonal covariance has lower accuracy in terms of the posterior standard deviations.
For the ionosphere and mushroom data,
DAIS is most accurate.
VI with full covariance performs well too while AMIS and VI with a diagonal covariance result in notable estimation error including underestimation of posterior standard deviations.

\begin{figure}
\begin{center}
\includegraphics[width=\textwidth]{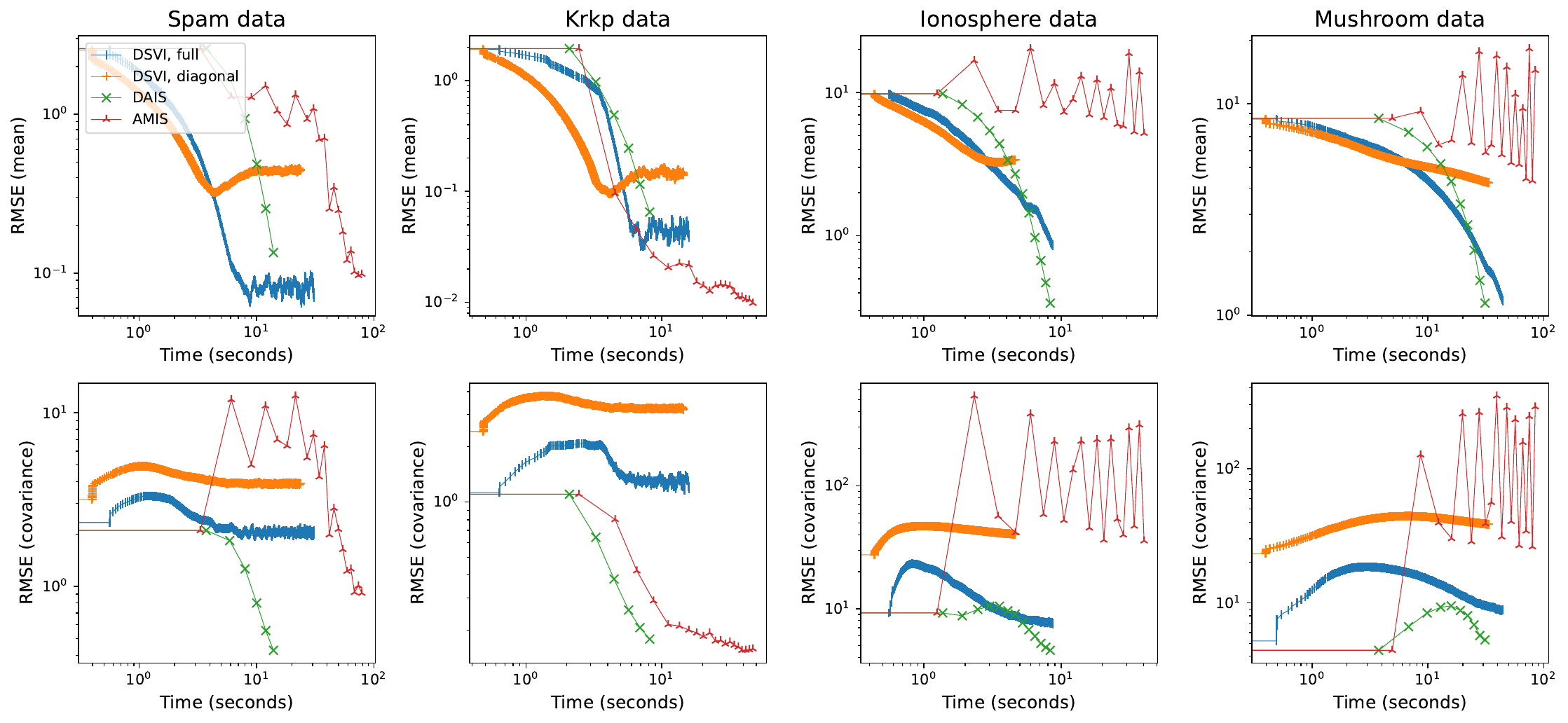}
\end{center}
\caption{
Approximation accuracy across iterations versus computation time. The accuracy is characterized by the RMSE quantities $\|\widehat{\mu}_{t} - \mu\|$ and ${\| \widehat{\Gamma}_{t} - \Gamma \|_{\mathrm{F}}}$, where
$\mu$ (resp.\ $\Gamma$) is the posterior mean (resp.\ covariance) from HMC,
$\mu_t$ and $\Gamma_t$ are the estimates at the $t$th iteration of DSVI/DAIS/AMIS,
and
$\|M\|_{\mathrm{F}} = (\sum M_{i,j}^2)^{1/2}$ is the Frobenius norm of the matrix $M$.
\label{fig:uci_times}}
\end{figure}

Figure~\ref{fig:uci_times} summarizes the computation times of the various algorithms. We used a computer with 8 CPU cores and 32 GB of RAM, specifically the machine type \texttt{e2-standard-2} from Google Cloud Platform.
The results are mixed across the four datasets with DAIS adaptive stopping rule resulting in a trade-off between computation time and accuracy that is competitive. We note that computation times are highly dependent on implementation and tuning parameters such as number of particles, and Figure~\ref{fig:uci_times} should be considered with this limitation in mind.

\section{Synthetic Inverse Problem}
\label{ap:inverse}

We consider the inverse problem from \citet{vandenBoom2019} as a synthetic problem that is higher-dimensional than the two-dimensional examples in Section~\ref{sec:2D}.
Consider the function $f:\mathbb{R}\to\mathbb{R}$ distributed as a zero-mean Gaussian process with covariance function $k(t, t') = \exp\{ - 400\, (t-t')^2 \}$. Then, $x \equiv \{ f\!\left(\frac{t-1}{d-1}\right),\, t = 1, \dots, d = 100 \}^\top$
is a discretization of $f$.
Define the $d\times d$ blurring matrix $G$ by first setting $G_{ij} = \exp\{ -\min(i+j,\, d-i-j)^2/25 \}$ for $1\leq i,j\leq d$ and then scaling its rows to sum to one.
Consider the $d$-dimensional vector $\overline{H}(x) = G\, x^{\odot 3}$ obtained by multiplying the matrix $G$ by the vector $x^{\odot 3} = \{x_i^3,\, i=1,\dots,d\}^\top$.
Then, generate a 30-dimensional vector $H(x)$ by sampling $30$ elements at random with replacement from the elements of $\overline{H}(x)$ with odd indices, as a type of subsampling. Finally, we generate $30$-dimensional data according to $y \sim \Normal\{ H(x), \mathrm{I}_{30} \}$ with $x$ fixed to a prior draw.
The target density follows as the posterior defined by this likelihood and the Gaussian prior $p_0$ induced by the Gaussian process, $\pi(x) \propto p_0(x)\, \Normal\{y \!\mid\! H(x), I_{30} \}$.
Computing $\pi$ is a Bayesian inverse problem where $H$ is the forward map that maps $x$ to a distribution on $y$. The goal is to ``invert'' $H$ by inferring $x$ from $y$.

We compare three Gaussian approximations of $\pi$. The first is the proposed Algorithm~\ref{alg:DAIS} with importance sample size $S=10^4$, $N_{\ess}=100$
and robustness parameter $c=1$
with which DAIS finishes in 20 iterations with $\gamma_t = 0.30$.\footnote{Here, we use the stopping criterion detailed in \citet{vandenBoom2024}.}
The second is a Laplace approximation from \citet{Steinberg2014} based on a Taylor series linearization of the forward map $H$.
Lastly, we run EP-IS with covariance matrix tapering from \citet{vandenBoom2019}, which exploits that the data are independent. As an arbitrarily accurate MC baseline, we run a preconditioned Crank–Nicolson algorithm \citep{Cotter2013} with the prior $p_0$ as reference measure for 100,000 iterations, which is substantially slower than the Gaussian approximations.

\begin{figure}
\begin{center}
\includegraphics{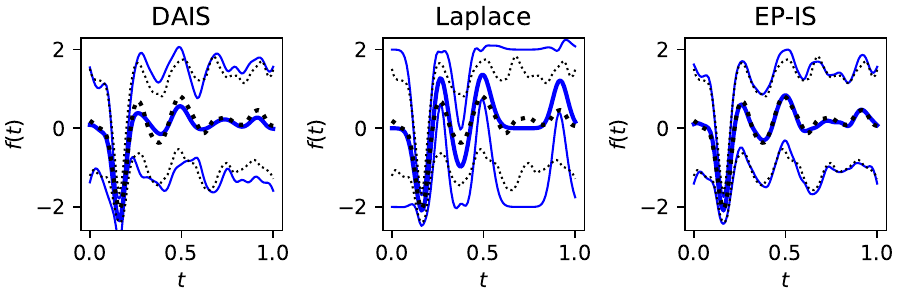}
\end{center}
\caption{
Posterior mean (thick line) and 2.5\% and 97.5\% quantiles (thin lines) of $\pi$ from the inverse problem as estimated by the preconditioned Crank–Nicolson algorithm (dotted),
and compared with estimates from DAIS, the Laplace approximation and EP-IS (solid).
\label{fig:inverse_problem}}
\end{figure}

Figure~\ref{fig:inverse_problem} summarizes the results of the different approximations.
The Laplace approximation is both the least accurate and the fastest approximation, taking only 1.0 seconds. DAIS and EP-IS are similarly accurate.
DAIS is faster than EP-IS (13 versus 16 seconds). Thus, DAIS is faster than the approximation that exploits the structure of the inverse problem without any problem-specific adjustments or reduced approximation accuracy.

\begin{figure}
\begin{center}
\includegraphics[width=\textwidth]{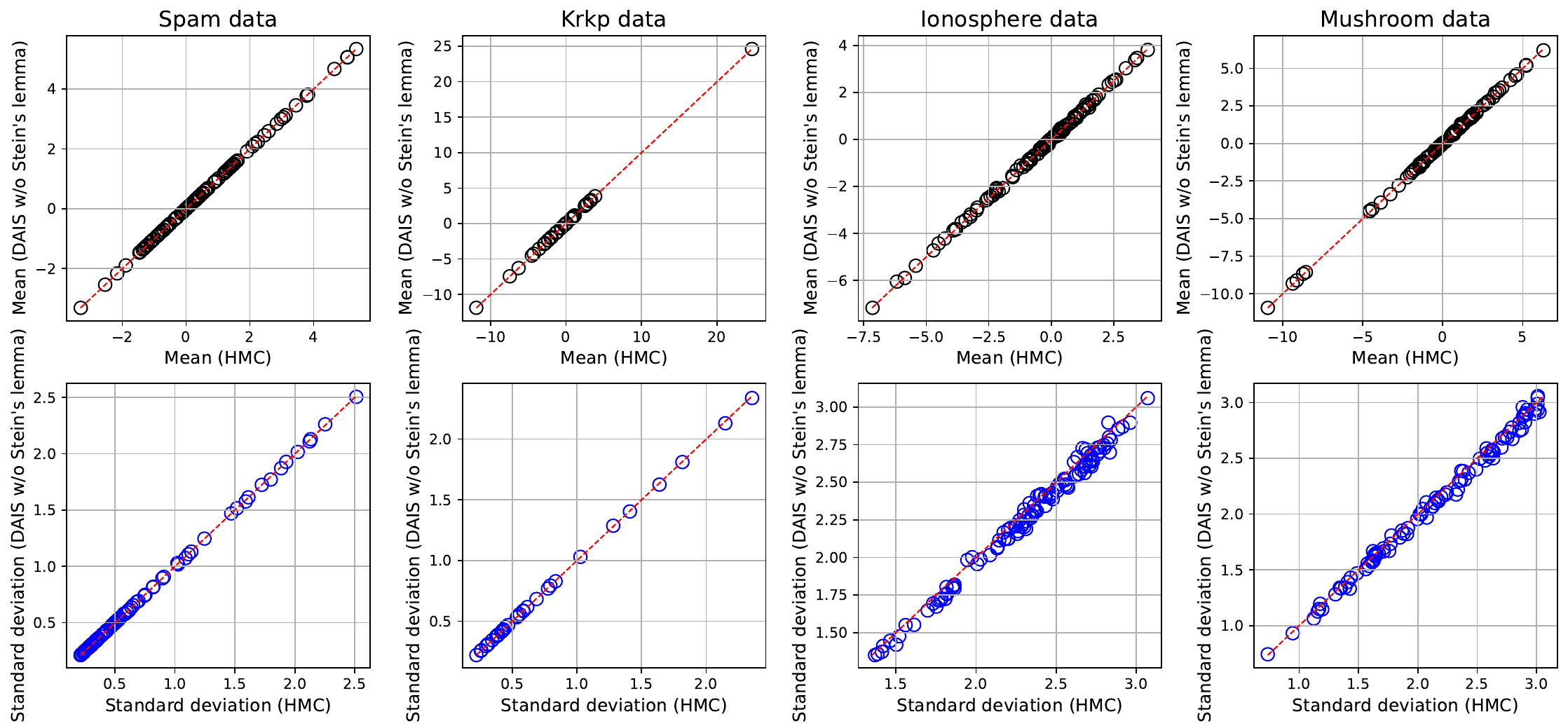}
\end{center}
\caption{
Scatter plots of the estimates from DAIS without using
Stein's identity to reduce MC error
through the identities in \eqref{eq.CV.mean.cov} (see Section~\ref{sec.CV}) versus the Hamiltonian MC estimates of the posterior means and standard deviations for the logistic regression examples.
\label{fig:uci_no_Stein}}
\end{figure}

\FloatBarrier
\bibliographystyle{chicago_no_eds_no_month}  
\bibliography{DAIS}